\documentclass[a4paper,reprint,floatfix]{revtex4-1}
\usepackage{graphicx}
\usepackage{float}
\usepackage{amsmath,amssymb,amsfonts}
\usepackage{array}
\usepackage{xcolor}
\usepackage{tabularx}

\graphicspath{{./}{figures/}}

\usepackage[caption=false]{subfig}

\newcommand{\intR}{\int_{\mathbb R}}
\newcommand{\intS}{\int_{\mathbb S}}
\newcommand{\ii}{\mathrm{i}}

\newtheorem{remark}{Remark}

\begin{document}

\title{Self-consistent Method and Steady States of Second-order Oscillators}

\author{Jian Gao}
\email[E-mail:]{jian.gao@rug.nl}
\affiliation{Bernoulli Institute for Mathematics, Computer Science, and Artificial Intelligence, University of Groningen, P.O. Box 407, 9700 AK, Groningen, The Netherlands}

\author{Konstantinos Efstathiou}
\email[E-mail:]{k.efstathiou@rug.nl}
\affiliation{Bernoulli Institute for Mathematics, Computer Science, and Artificial Intelligence, University of Groningen, P.O. Box 407, 9700 AK, Groningen, The Netherlands}

\begin{abstract}
The self-consistent method, first introduced by Kuramoto, is a powerful tool for the analysis of the steady states of coupled oscillator networks.
For second-order oscillator networks complications to the application of the self-consistent method arise because of the \emph{bistable} behavior due to the co-existence of a stable fixed point and a stable limit cycle, and the resulting complicated boundary between the corresponding basins of attraction.
In this paper, we report on a self-consistent analysis of second-order oscillators which is simpler compared to previous approaches while giving more accurate results in the small inertia regime and close to incoherence.
We apply the method to analyze the steady states of coupled second-order oscillators and we introduce the concepts of \emph{margin region} and \emph{scaled inertia}.
The improved accuracy of the self-consistent method close to incoherence leads to an accurate estimate of the critical coupling corresponding to transitions from incoherence.
\end{abstract}

\maketitle

\section{Introduction}

Synchronization of coupled dynamical units has been recognized in the past 50 years, since the pioneering works of Winfree \cite{Winfree1967} and Kuramoto \cite{Kuramoto1975}, as one of the most important phenomena in nature.
Several mathematical models are used to understand this fascinating phenomenon.
Among them, coupled Kuramoto oscillators is one of the most popular models \cite{Kuramoto1987, Rodrigues2016}.
Many analytical methods have been developed for Kuramoto oscillators, like the self-consistent method \cite{Sakaguchi1986, Kuramoto1987}, the Ott-Antonsen ansatz \cite{Ott2008, Pikovsky2011, Marvel2009}, and stability analysis in the continuum limit \cite{Strogatz1991, Crawford1994, Chiba2015}.
With these methods, complemented by numerical simulations, many interesting phenomena of Kuramoto oscillators have been found and analyzed \cite{Strogatz2000, Acebron2005, Rodrigues2016, Arenas2008}.

The Kuramoto model is not only simple and amenable to analytical considerations, but it is also easy to generalize in different directions.
By adding frequency adaptations (inertias), the second-order oscillators model has been proposed and developed to describe the dynamics of several systems:
tropical Asian species of fireflies \cite{Ermentrout1991};
Josephson junction arrays \cite{Levi1978, Watanabe1994, Trees2005};
goods markets \cite{Ikeda2012};
dendritic neurons \cite{Sakyte2011};
and power grids \cite{Filatrella2008}.
Many important conclusions about the stability of power grids have been obtained through analysis of this model \cite{Rohden2012, Rohden2014, Lozano2012, Witthaut2012, Menck2013, Hellmann2016, Kim2015, Gambuzza2017, Dorfler2013c, Grzybowski2016, Maizi2016, Manik2016a, Pinto2016, Rohden2017, Witthaut2016}. 

Kuramoto's self-consistent analysis \cite{Kuramoto1975} has been extended to second-order oscillators by Tanaka \emph{et al} \cite{Tanaka1997, Tanaka1997a}.
In this paper, we are revisiting the self-consistent method for the steady states of second-order oscillators.
The benefits are twofold. 
First, we considerably simplify the derivation of the estimates of the limit cycles of the system that play a role in the self-consistent analysis. 
Second, the obtained estimates are much more accurate compared to earlier estimates, especially, for small inertias.
Therefore, the method can be applied to the general case of the second-order oscillators, with large or small inertias, for both incoherent or synchronized states.
The improved limit cycle estimates also lead to self-consistent equations that coincide well with numerical simulations.
Moreover, the more accurate self-consistent method allows us to obtain the critical coupling strength $K_c$ where steady state solutions bifurcate from the incoherent state. 
The results agree with the stability analysis of the incoherent state in \cite{Barre2016} obtained through an unstable manifold expansion of the associated continuity equation.

We give a short outline of the paper. 
In Sec.~\ref{section_model}, the model and the general framework of the self-consistent method are introduced.
The dynamics of a single second-order oscillator is discussed in Sec.~\ref{section_single}.
Based on this, the self-consistent equation is obtained in Sec.~\ref{section_self} and several properties of steady states for arbitrary natural frequency distributions are discussed.
In Sec.~\ref{section_symmetry} the steady states of oscillators with symmetric and unimodal natural frequency distribution are discussed and the theoretical results are compared to numerical simulations.
We conclude the paper in Sec.~\ref{section_conclusion}.

\section{Model and self-consistent method}\label{section_model}

The model for coupled second-order Kuramoto-type oscillators reads
\begin{equation}\label{Dynamic-of-single}
m_i \ddot{\varphi}_i + D_i \dot{\varphi}_i
= \Omega_i + \frac{K}{N} \sum_{j=1}^{N} \sin(\varphi_j-\varphi_i), 
\end{equation}
for $i=1,\dots,N$, where $m_i$, $D_i$, and $\Omega_i$ are respectively the inertia, damping coefficient, and natural frequency of the $i$-th oscillator. The dynamics of each oscillator is described by its phase $\varphi$ and corresponding velocity $\dot\varphi$, with $(\varphi,\dot\varphi) \in \mathbb S \times \mathbb R$, where $\mathbb S = \mathbb R / 2\pi \mathbb Z$ (a circle of length $2\pi$). Moreover, $N$ is the number of oscillators and $K$ is the (uniform) coupling strength.

To describe the collective behavior of the oscillators, one defines the order parameter as
\begin{equation}\label{order_definition_eq}
  r e^{\ii\phi} = \frac{1}{N}\sum_{j=1}^{N}e^{\ii\varphi_j}.
\end{equation}
Here $r \in [0,1]$ is indicative of the \emph{coherence} of the oscillators.
We have $r=1$ if and only if all the oscillators are synchronized with $\varphi_i(t)=\varphi(t)$ for all $1\leq i\leq N$.
Moreover, an (almost) uniform distribution of the phase terms $e^{\ii \varphi_i}$ over the unit circle corresponds to $r \approxeq 0$; note, however, that the opposite implication is not always true.
The rate of change of the collective phase $\phi$ is related to the mean frequency of the oscillators and describes the global rotation.

Using the amplitude $r(t)$ and phase $\phi(t)$ of the order parameter, 
the model \eqref{Dynamic-of-single} can be rewritten in a mean-field form as
\begin{equation}\label{single_mean_eq}
  m_i \ddot{\varphi}_i + D_i \dot{\varphi}_i
  = \Omega_i + Kr(t)\sin(\phi(t)-\varphi_i),
\end{equation}
with $i=1,\dots,N$.
In this paper we consider only \emph{steady states}, given by
\begin{equation}\label{def_steady_eq}
  r(t) = r, \quad \phi(t)=\Omega^r t + \Psi,
\end{equation}
where $r$, $\Omega^r$, and $\Psi$ are all constant. 
Passing to a frame rotating by $\phi(t) = \Omega^r t + \Psi$, and defining the phase difference between each oscillator and the frame as
\begin{equation}
  \theta_i = \varphi_i -\phi(t),
\end{equation}
one finds that the dynamics for the oscillators in the rotating frame is given by
\begin{equation}\label{phase_differece_eq}
  m_i \ddot{\theta}_i + D_i \dot{\theta}_i
  = (\Omega_i - D_i \Omega^r) - K r \sin\theta_i,
\end{equation}
for $i=1,\dots,N$. 
Dropping the index $i$ from Eq.~\eqref{phase_differece_eq} and assuming $Krm \ne 0$, the dynamics of a single oscillator in the rotating frame can be transformed to the standard form 
\begin{equation}\label{single_dynamic_eq}
\ddot{\theta}+a\dot{\theta} = b - \sin\theta,
\end{equation}
with only two effective parameters
\begin{equation}
a=\frac{D}{\sqrt{Krm} }, \quad  b=\frac{\Omega-D\Omega^r}{Kr},
\end{equation}
and rescaled time $\tau=t\sqrt{Kr/m}$.

In this paper we consider only the case where all the oscillators have the same inertia $m$ and damping coefficient $D$ even though our approach generalizes to the case of different inertias and damping coefficients. 
To pass to the continuum limit we replace by a density function $g(\Omega,\theta_0,\dot\theta_0)$ the collection of discrete oscillators characterized by natural frequency $\Omega_i$ and initial state $(\theta_i(0), \dot{\theta}_i(0))$. 
Note that, differently from the case of first-order Kuramoto oscillators, the initial state $(\theta_0,\dot\theta_0)$, is important for the dynamics of our case because of the bistable mechanism we discuss in Sec.~\ref{section_single}.

In terms of the phases $\theta_i$ Eq.~\eqref{order_definition_eq} becomes
\begin{align*}
r = \frac{1}{N} \sum_{j=1}^N e^{\ii \theta_j},  
\end{align*}
which in the continuum limit reads as
\begin{equation}
\begin{aligned}\label{self_consistent_eq_continue}
r &= \intS \intR \intR
g(\Omega,\theta_0,\dot\theta_0) e^{\ii \theta(t)}
\, d\theta_0\, d\dot\theta_0\, d\Omega.
\end{aligned}
\end{equation}
Here $\theta(t)$ represents the solution for the dynamics of a single oscillator, Eq.~\eqref{phase_differece_eq}, and depends on parameters $(a,b)$ and initial conditions $(\theta_0,\dot\theta_0)$.
In what follows, our goal is to understand the self-consistent equation~\eqref{self_consistent_eq_continue} and use it to explore the properties of steady states of second-order oscillators.

\section{Dynamics of a single oscillator and bistable region}
\label{section_single}

In this section we recall facts about the dynamics of a single second-order oscillator described by Eq.~\eqref{phase_differece_eq} and then we compute an approximation to the limit cycle that plays a central role in what follows.

\begin{figure}
  \centering
  \includegraphics[scale=0.40]{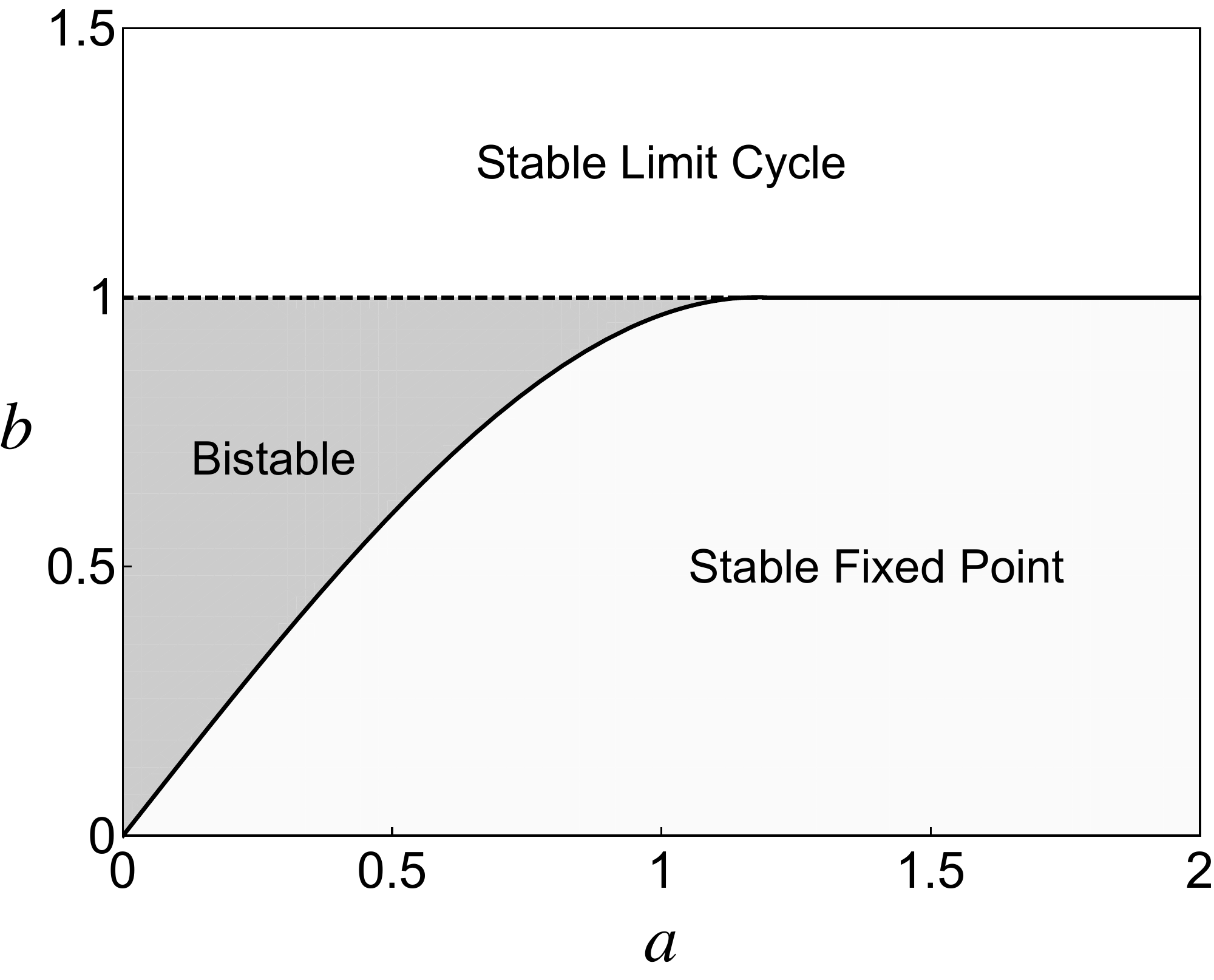}
  \caption{Phase diagram for a single second order oscillator, Eq.~\eqref{single_dynamic_eq}, in the $(a,b)$ parameter plane. The thick black curve represents $b_S(a)$, and the dashed curve represents $b_L(a) = 1$. Note that for $a \ge 1.193$ the two curves coincide. The two curves separate the parameter plane into the three regions shown in the diagram.}
  \label{fig_phase}
\end{figure}

\subsection{Fixed points and limit cycle}

\begin{figure*}
  \centering
  \includegraphics[width=\textwidth]{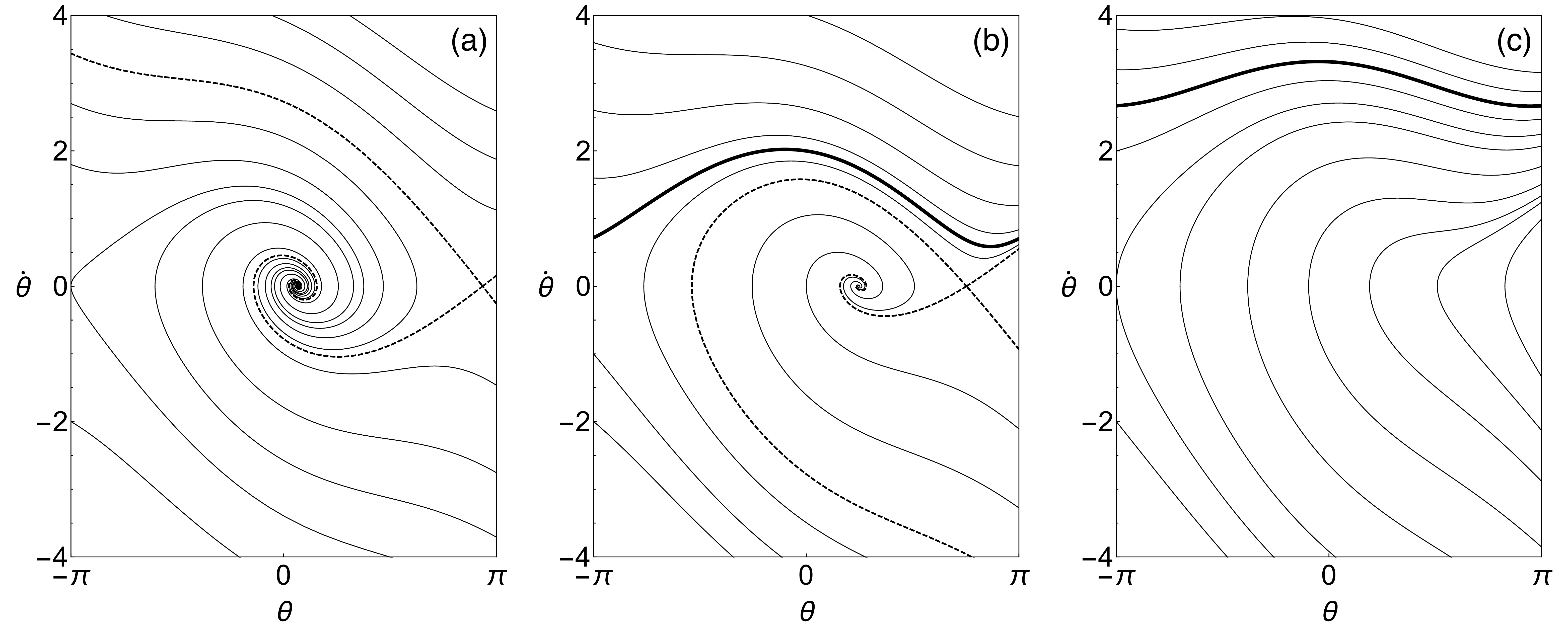}
  \caption{Phase portraits for a single second order oscillator, Eq.~\eqref{single_dynamic_eq}. In (a), for $a=0.5$, $b=0.2$, there is a stable and an unstable fixed point and no limit cycles. In (b), for $a=0.5$, $b=0.7$, we have a bistable system where the stable and unstable fixed points co-exist with a limit cycle. In (c), for $a=0.5$, $b=1.5$, there exists only a limit cycle. The solid thick curve in figures (b) and (c) represents the limit cycle. The dashed curves in figures (a) and (b) represent stable and unstable asymptotic curves to the saddle fixed point. Note that the left and right sides of the picture must be identified because of the $2\pi$-periodicity of $\theta$.}
  \label{fig_phase_portraits_single}
\end{figure*}

Depending on the values of the parameters $a$ and $b$, Eq.~\eqref{single_dynamic_eq} can have either a fixed point, or a globally attracting stable limit cycle, or a bistable region where the fixed point and the limit cycle coexist \cite{Levi1978, Tanaka1997a, Strogatz2014}.
A thorough qualitative study of the fixed points and limit cycle in this system can be found in \cite{Levi1978} where it has been shown that for $b > 1$ the system has no fixed points and it has a globally attracting stable limit cycle.
For $b < 1$ the system has exactly two fixed points, one stable and one unstable.
Then for each fixed value $b < 1$ there is a value $a_*(b)$ of $a$ for which if $0 < a < a_*(b)$ the system also has a stable limit cycle; a so-called \emph{bistable} state.
For $a > a_*(b)$ the limit cycle does not exist anymore.
The transition at $a = a_*(b)$ occurs through a homoclinic tangency bifurcation. For $b = 1$ the situation concerning the limit cycle is similar, with corresponding $a_*(1) \approxeq 1.193$. However, for $b=1$ the two fixed points merge and the system undergoes a saddle-node bifurcation so that for $b > 1$ there are no more fixed points. These results are summarized in the phase diagram shown in Fig.~\ref{fig_phase}. The dynamics for three qualitatively different cases are shown in Fig.~\ref{fig_phase_portraits_single}.

\begin{remark}
  The limit cycle is a \emph{running} or \emph{rotating} limit cycle. That is, following the dynamics on the limit cycle, in one period the phase $\theta$ increases by $2\pi$. Alternatively stated, the phase space of the system is a cylinder $(\mathbb R / 2 \pi \mathbb Z) \times \mathbb R$ and the limit cycle is a non-homotopically-trivial circle on the cylinder, see Fig.~\ref{fig_phase_portraits_single}(b) and Fig.~\ref{fig_phase_portraits_single}(c). 
\end{remark}

For our purposes we use a different description of the phase diagram. We define two functions: the (constant) function $b_L(a) = 1$, and the function
\begin{align*}
  b_S(a) = \begin{cases}
    (a_*)^{-1}(a), & 0 \le a \le a_*(1) \approxeq 1.193, \\
    1, & a \ge a_*(1),
  \end{cases}
\end{align*}
where $(a_*)^{-1}$ is the inverse of $a_* : [0,1] \to [0,a_*(1)]$. Clearly, adapting the discussion above, for $b > b_L(a) = 1$ there exists a globally attracting limit cycle and no fixed points. For $0 < b < b_S(a)$ the system has two fixed points and no limit cycle. Finally, the bistable state exists for $b_S(a) < b < b_L(a)$.

The condition $b < b_L(a) = 1$ for the existence of fixed points is easily obtained, since the fixed points correspond to solutions of $(\dot\theta, \ddot\theta) = (0,0)$, giving the equation $b = \sin\theta$. Further computing the stability of the fixed points we obtain that for $b < b_L(a) = 1$ (and $a > 0$) the system has the fixed points
\begin{equation}
  \begin{aligned}
    (\theta_0 ,\, \dot\theta_0) &= (\arcsin(b) ,\, 0), \qquad && \text{(stable)}, \\
    (\theta_0 ,\, \dot\theta_0) &= (\pi-\arcsin(b) ,\, 0), \qquad && \text{(saddle point)}.
  \end{aligned}
\end{equation}
In particular, for the stable point we obtain 
\begin{equation}\label{eq_zl}
\exp(\ii \theta_0)=\sqrt{1-b^2} + \ii b.
\end{equation} 

Determining the existence region of the limit cycle, that is, the function $b_S(a)$, is more complicated. The limit cycle is always stable and it appears for $0 < a < a_*(1)$ through a homoclinic bifurcation and for $a > a_*(1)$ through an infinite period bifurcation \cite{Levi1978, Strogatz2014}. For small values of $a$ an application of Melnikov's method \cite{Guckenheimer2013}, or Lyapunov's direct method \cite{Risken1996fokker}, gives
\begin{equation}
  b_S(a) \approxeq 4 \pi^{-1} a \approxeq 1.2732\,a,
\end{equation}
cf.\ Fig.~\ref{fig_phase}.
Using numerical simulations, see \cite{Belykh2016}, a higher order approximation of this bifurcation line has been obtained as
\begin{equation}
b_S(a) \approxeq
  \begin{cases}
    1.2732\,a - 0.3056\,a^3,  & 0 \le a \le a_*(1) \approxeq 1.193, \\
    1, & a \ge a_*(1).
\end{cases}
\end{equation}

\subsection{Approximation of the limit cycle}

The analysis of the self-consistent equation for the second-order oscillators requires an analytic expression for the limit cycle.
In general, the solution of the limit cycle cannot be obtained analytically.
An approximate expression has been computed in \cite{Tanaka1997a} through the use of the Poincar\'e-Lindstedt method at the \emph{underdamped} limit $a^2 \ll 1 \sim b$.
Translating the result of \cite{Tanaka1997a} to our notation we have
\begin{subequations}\label{eq_tanaka_approx}
\begin{equation}
  \theta(\tau) = \nu\tau + \frac{a^2}{b^2} \sin(\nu\tau) 
  + \frac{a^4}{b^3} (\cos(\nu\tau)-1) + \cdots,
\end{equation}
where
\begin{equation}
  \nu = \frac{b}{a} - \frac{a^3}{2b^3} + \cdots.
\end{equation}
\end{subequations}
The value of the time-average of $\cos\theta$ on the limit cycle is then approximated in \cite{Tanaka1997a} by
\begin{equation}\label{eq_avg_cos_tanaka}
  \langle \cos\theta \rangle = - \frac{a^2}{2b^2},
\end{equation}
see also \footnote{In Eq.~(A.3) of \cite{Tanaka1997a} the values of $\cos C$ and $\sin C$ have been interchanged leading to an incorrect estimation of the time-average of $\cos\theta$ along the limit cycle. In particular, the expression $\frac12 \tilde{r} \Delta^3$ in Eq.~(34) of \cite{Tanaka1997a} should have been $\frac12 \tilde{r} \Delta^2$ which, in our notation, corresponds to Eq.~\eqref{eq_avg_cos_tanaka} in the present paper.}.

Here we derive different approximations to the limit cycle and the corresponding value of $\langle \cos\theta \rangle$ which are valid for a larger range of parameter values and which at the underdamped limit coincide with Tanaka's approximations, Eq.~\eqref{eq_tanaka_approx} and Eq.~\eqref{eq_avg_cos_tanaka}.

We start by expressing $\dot\theta$ as a function of $\theta$ for points on the limit cycle using a Fourier series. Keeping only the first harmonics we write
\begin{align*}
  \dot\theta(\theta) = A_0 + A_1 \cos\theta + B_1 \sin\theta.
\end{align*}
Substituting the last expression in Eq.~\eqref{single_dynamic_eq} and computing the Fourier coefficients so that the first harmonics vanish we obtain
\begin{align}\label{eq_dottheta_theta}
\begin{aligned}
  \dot\theta(\theta)
  & = \frac{b}{a} + \frac{a b}{a^4 + b^2} \cos\theta - \frac{a^3}{a^4+b^2} \sin\theta \\
  & = \nu_0 + \varepsilon \cos(\theta+\theta_*),
\end{aligned}
\end{align}
where
\begin{align*}
  \nu_0 = \frac{b}{a}, \quad \frac{1}{\varepsilon} e^{\ii \theta_*} = \nu_0 + i a.
\end{align*}

The time-average of $e^{\ii\theta}$ on the limit cycle is given by
\begin{equation}
\begin{aligned}
  \langle e^{\ii\theta} \rangle
  & = \frac{1}{T} \int_0^T \exp(\ii\theta(\tau)) \, d\tau \\
  & = \int_0^{2\pi} \frac{\exp(\ii\theta)}{\dot\theta(\theta)} \, d\theta 
    \bigg/ 
    \int_0^{2\pi} \frac{1}{\dot\theta(\theta)} \, d\theta.
\end{aligned}
\end{equation}
These integrals can be exactly computed for $\dot\theta(\theta)$ given by Eq.~\eqref{eq_dottheta_theta}.
Computing the period integral we obtain
\begin{equation}\label{eq_approx_freq}
  \nu 
  = \frac{2\pi}{T} = \sqrt{\nu_0^2 - \varepsilon^2} 
  = \sqrt{\nu_0^2 - (\nu_0^2+a^2)^{-1}}
  \le \nu_0.
\end{equation}
The computation of $\langle e^{\ii\theta} \rangle$ gives
\begin{equation}\label{eq_avg_exp}
\begin{aligned}
  \langle e^{\ii\theta} \rangle 
  & = e^{-\ii\theta_*} \varepsilon^{-1} \Big[\sqrt{\nu_0^2-\varepsilon^2}-\nu_0 \Big] \\
  & = - \nu_0 (\nu_0-\nu) + \ii\, a (\nu_0-\nu).
\end{aligned}
\end{equation}
A Taylor series expansion in $\varepsilon \ll 1$ gives the expression
\begin{equation}
\begin{aligned}
  \langle e^{\ii\theta} \rangle 
  & = \frac12 \Big( -1 + \frac{\ii a}{\nu_0} \Big) \varepsilon^2 + O(\varepsilon^4) \\
  & = \frac12 \Big( -1 + \frac{\ii a^2}{b} \Big) \frac{a^2}{a^4+b^2} + O(\varepsilon^4).
\end{aligned}
\label{eq_used_taylor}
\end{equation}
which is valid for $a^2 + \nu_0^2 \gg 1$, that is, for $a^2 \ll b$ or $a \gg 1$. With the same order of approximation, one can replace $\theta$ by $\nu_0 \tau$ in Eq.~\eqref{eq_dottheta_theta}.
Then integration with respect to $\tau$ gives
\begin{equation}\label{eq_approx_new}
\theta(\tau) = \nu_0\tau + \frac{a^4}{b (a^4 + b^2)} \big[ \cos(\nu_0\tau) - 1 \big]
	+ \frac{a^2}{a^4 + b^2} \sin(\nu_0\tau),
\end{equation}
with the constant of integration chosen so that $\theta(0) = 0$. 
Observe that for $a^2 \ll b$, Eq.~\eqref{eq_approx_new} gives the approximation in Eq.~\eqref{eq_tanaka_approx}, and the real part of Eq.~\eqref{eq_used_taylor} gives the approximation in Eq.~\eqref{eq_avg_cos_tanaka}.
	
\begin{figure*}
  \centering
  \includegraphics[width=1.0\textwidth]{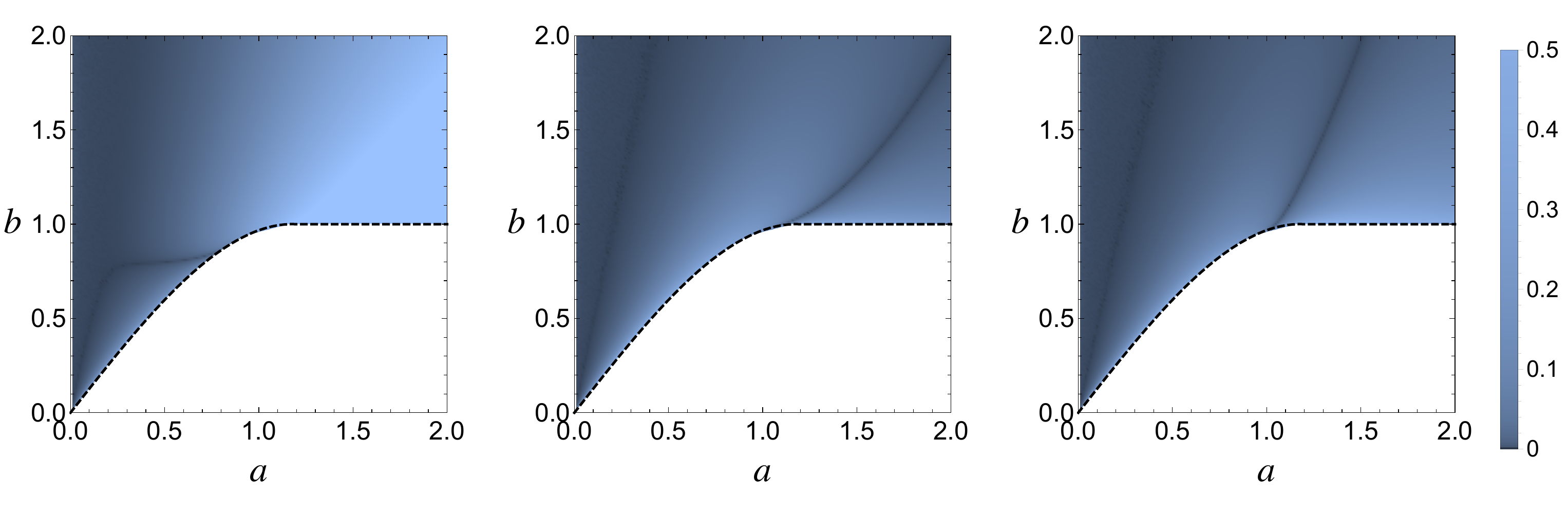}
  \caption{Approximation errors of $\langle \cos\theta\rangle$ corresponding from left to right to Eq.~\eqref{eq_avg_cos_tanaka}, Eq.~\eqref{eq_used_taylor}, and Eq.~\eqref{eq_avg_exp}. 
  The error is defined as the maximum of the absolute value of the difference between the numerical calculation of $\langle \cos\theta \rangle$ along the limit cycle (which exists only for $b > b_S(a)$) and the corresponding analytical estimation. 
  Note that in panel (a) all errors above $0.5$ are represented by the same color.
  }
  \label{fig_approximation_limit_cycle}
\end{figure*}

As a result, when $a^2 \ll 1 \sim b$ all the three approximate expressions, Eq.~\eqref{eq_avg_exp}, Eq.~\eqref{eq_used_taylor}, and Eq.~\eqref{eq_avg_cos_tanaka}, give the same estimation of $\langle \cos\theta \rangle$ on the limit cycle.
Using numerical simulations, we have found that both the computation in Eq.~\eqref{eq_avg_exp} or the one with the Taylor expansion in Eq.~\eqref{eq_used_taylor} are significantly better estimates of $\langle \cos\theta \rangle$ on the limit cycle compared to the approximation obtained previously as Eq.~\eqref{eq_avg_cos_tanaka}, see Fig.~\ref{fig_approximation_limit_cycle}. 
However, it is hard to distinguish from these numerical results which one of Eq.~\eqref{eq_approx_freq} or Eq.~\eqref{eq_approx_new} provides the best approximation.
Moreover, in the limit of large or small inertias, with $a^2 \ll b$ or $a \gg 1$, Eq.~\eqref{eq_avg_exp} and Eq.~\eqref{eq_used_taylor} are the same neglecting terms of order $O(\varepsilon^4)$ and higher.
Hence we consider both expressions as equally accurate for the self-consistent method. 
In the computations in subsequent sections we will be using the expression Eq.~\eqref{eq_used_taylor} because it leads to simpler analytical expressions.
One can show that both the quantitative (such as the value of $K_c$) and qualitative (such as the margin regions) results we obtain can also be obtained with the alternative expression, Eq.~\eqref{eq_avg_exp}.

\section{Self-consistent equation for two processes}\label{section_self}

Because of the complexity of the basins of attraction, it is difficult to study the problem of synchronization in its full generality.
Instead, following Tanaka \emph{et al}'s approach in \cite{Tanaka1997, Tanaka1997a}, we consider the synchronization during the so-called \emph{forward} and \emph{backward} processes.

In the forward process $[F]$ the system starts at the incoherent state with coupling $K=0$ and then $K$ progressively increases.
Small coupling $K \ll 1$ and incoherent state $r \approxeq 0$ corresponds to large values of $a^2=D^2/Krm$ and $b=(\Omega-D\Omega^r)/Kr$.
In particular, it can be ensured that all oscillators are in the parameter region $b > b_L(a) = 1$ where there exists only a stable limit cycle (no stable fixed point), see Sec.~\ref{section_single}.
Similarly, in the backward process $[B]$ the system starts at a coherent state with a large value of $K$ and then the coupling progressively decreases.
Large coupling $K \gg 1$ and coherent state $r \approxeq 1$ corresponds to small values of $a^2 = D^2/Krm$ and $b=(\Omega-D\Omega^r)/Kr$.
Here it can be ensured that all oscillators are in the parameter region $0 < b < b_S(a)$ where there exists only a stable fixed point (no stable limit cycle).

Thus, in these two processes the initial states of all the oscillators lie entirely in the basin of one stable state, the fixed point for $[B]$ and the limit cycle for $[F]$, and only leave it when this stable state disappears as $K$ changes.
This happens in the $[B]$ process when oscillators cross the boundary $b_P = b_L$ and in the $[F]$ process when they cross the boundary $b_P = b_S$.
Note that because of the different values of the natural frequency $\Omega$ for each oscillator, the oscillators will move from one stable state to another one at different values of $K$.

The previous discussion implies that in the forward and backward processes the role of the initial state $(\theta_0,\dot\theta_0)$ for each oscillator can be neglected and therefore we can consider the density $g(\Omega)$ obtained by integrating $g(\Omega,\theta_0,\dot\theta_0)$, that is,
\begin{align*}
  g(\Omega) = \intS \intR g(\Omega,\theta_0,\dot\theta_0) \, d\theta_0 \, d\dot\theta_0.
\end{align*}
Moreover, for fixed values of $r$ and $\Omega^r$ the density $g(\Omega)$ is transformed under the change of variables $b = (\Omega - D\Omega^r)/Kr$ to a density $G(b)$ as
\begin{align*}
  G(b) = Kr \, g(Krb+D\Omega^r).
\end{align*}

During the forward and backward processes, we can write the order parameter $r$ as the sum of the coherence of two populations of oscillators: oscillators at the stable fixed point, which we call \emph{locked}, and oscillators at the stable limit cycle, which we call \emph{running}.
We have
\begin{equation*}
  r = z_l + z_r,
\end{equation*}
where $z_l$ and $z_r$ represent the coherence of the locked and running oscillators, respectively. 
In the forward and backward processes, the locked and running oscillators are separated by the boundary of the bistable region of a single oscillator, i.e., by $b_S$ for $[F]$ and $b_L$ for $[B]$. 

With substitution of the stable solution of a single oscillator, Eq.~\eqref{eq_zl}, into the self-consistent equation for the locked oscillators, $z_l$ reads
\begin{align*}
  z_l
  & = \int_{|b| < b_P(a)} \,
    G(b) \left[\sqrt{1-b^2}+ib\right]
    db \\
  & = \intR G(b) \,
    \mathbf{1}_l \left[\sqrt{1-b^2}+ib\right]
    db,
\end{align*}
where the indicator function $\mathbf{1}_l$ takes the value $1$ if $|b| < b_P(a)$, corresponding to the condition for locked oscillators, and $0$ otherwise. 
In the equation above we have $b_P = b_L$ for the backward process and $b_P = b_S$ for the forward process.

For the running oscillators, using Eq.~\eqref{eq_used_taylor}, the coherence $z_r$ reads
\begin{align*}
  z_r
  & = \intR G(b) \,
  \mathbf{1}_r
  \langle e^{i\theta} \rangle
  db 
  \\
  & = \intR G(b) \,
  \mathbf{1}_r
  \left[ \frac12 \Big( -1 + \frac{\ii a^2}{b} \Big) \frac{a^2}{a^4+b^2}  \right]
  db,
\end{align*}
where the function $\mathbf{1}_r$ takes the value $1$ if $|b| > b_P(a)$, corresponding to the condition for running oscillators, and $0$ otherwise. 
Note that
\begin{align*}
  \mathbf{1}_l + \mathbf{1}_r = 0.
\end{align*}

Combining $z_l$ and $z_r$, and separating the real and imaginary parts, we obtain the self-consistent equations for the second-order oscillators as
\begin{subequations}
\label{self_consistent_eq_separate}
\begin{align}
r &= \intR
G(b) 
\left[ \mathbf{1}_l \sqrt{1-b^2} - \mathbf{1}_r \frac{a^2}{2(b^2+a^4)} \right]
db,
\label{eq_self_consistent_separate_real}
\\
0 &= \intR
G(b)
\left[ \mathbf{1}_l b + \mathbf{1}_r \frac{a^4}{2b(b^2+a^4)} \right]
db.
\label{eq_self_consistent_separate_imag}
\end{align}
\end{subequations}

One checks that Eq.~\eqref{self_consistent_eq_separate} always has the trivial solution $r=0$.
For $r>0$ with the definition $q=Kr$, we have $a = D / \sqrt{qm}$ and
Eq.~\eqref{self_consistent_eq_separate} can be divided by $q > 0$ to obtain the equations
\begin{widetext}
\begin{subequations}
\label{eq_self_consistent_calculation}
\begin{align}
  \frac{1}{K} &= F_1(q,\Omega^r)
  \equiv
  \intR g(qb+D\Omega^r) 
  \left[ \mathbf{1}_l \sqrt{1-b^2} - \mathbf{1}_r \frac{a^2}{2(b^2+a^4)} \right]
  db,
  \label{eq_self_consistent_calculation_real}
  \\
  0 &= F_2(q,\Omega^r)\equiv\intR
  g(qb+D\Omega^r) 
  \left[ \mathbf{1}_l b + \mathbf{1}_r \frac{a^4}{2b(b^2+a^4)}\right]
  db.
  \label{eq_self_consistent_calculation_imag}
  \end{align}
\end{subequations}
\end{widetext} 

We now describe how to solve the self-consistent equation~\eqref{eq_self_consistent_calculation}.
For each pair $(q,\Omega^r)$ satisfying $0=F_2(q,\Omega^r)$, one can obtain the corresponding value of $K$ by computing $1/K = F_1(q,\Omega^r)$, provided that $F_1(q,\Omega^r) > 0$. Since $q = K r$ we conclude that the triplet $(K,q,\Omega^r)$ can be transformed to the solutions of the self-consistent equation Eq.~\eqref{eq_self_consistent_calculation} as
\begin{align*}
  (K, r, \Omega^r)
  & = (K, q K^{-1}, \Omega^r) \\
  & = ([F_1(q,\Omega^r)]^{-1}, q F_1(q,\Omega^r), \Omega^r).
\end{align*}
These solutions can be (locally) parameterized in terms of $K$ as families $(r(K),\Omega^r(K))$, except at points of bifurcation, i.e., at values of $K$ where the number of families changes. 

\begin{figure*}
  \centering
  {\includegraphics[width=0.48\textwidth]{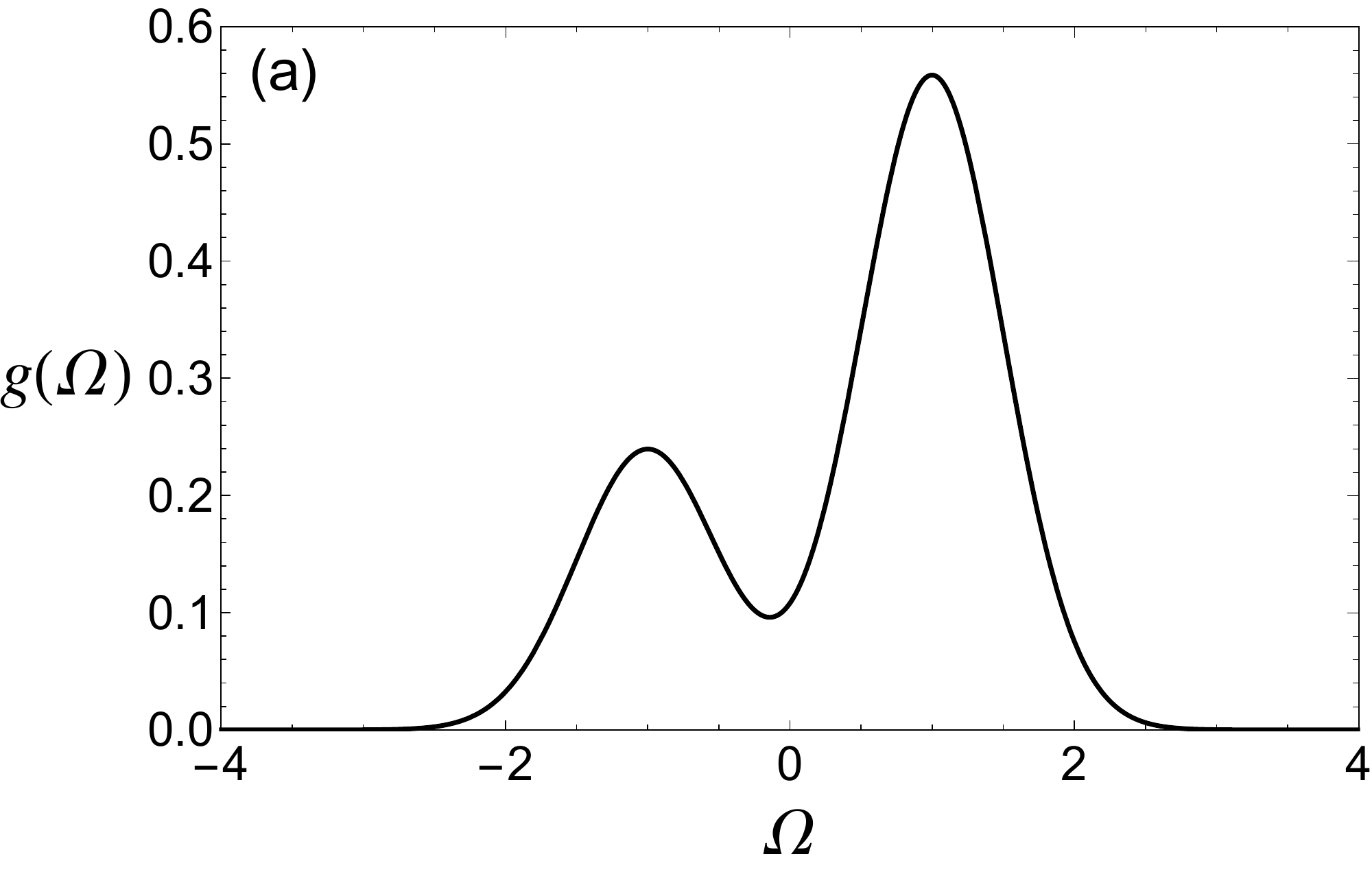}}
  \hfill
  {\includegraphics[width=0.48\textwidth]{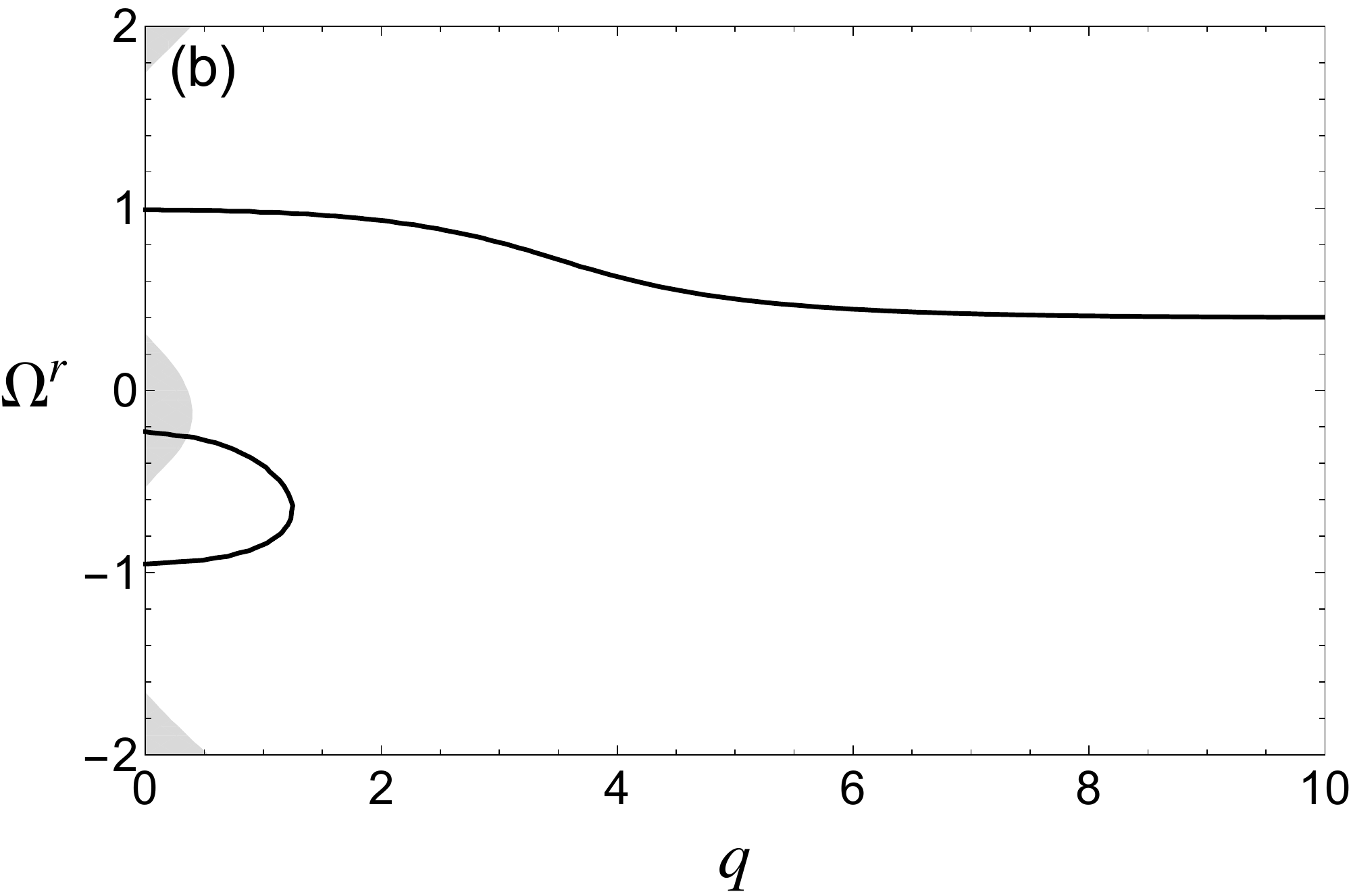}}
  \\
  {\parbox[b]{0.48\textwidth}{\centering\includegraphics[width=0.35\textwidth]{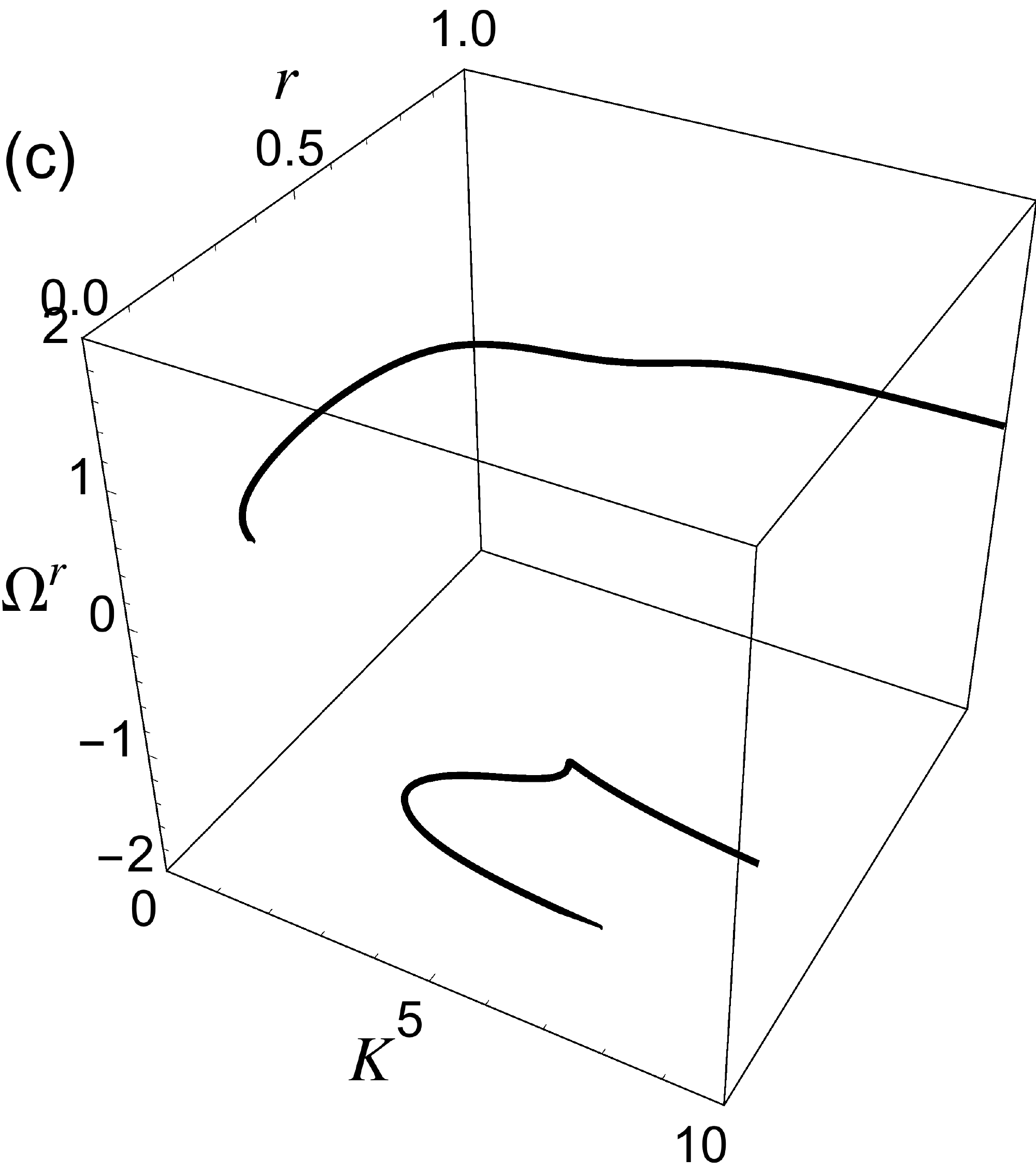}}}
  \hfill
  {\includegraphics[width=0.48\textwidth]{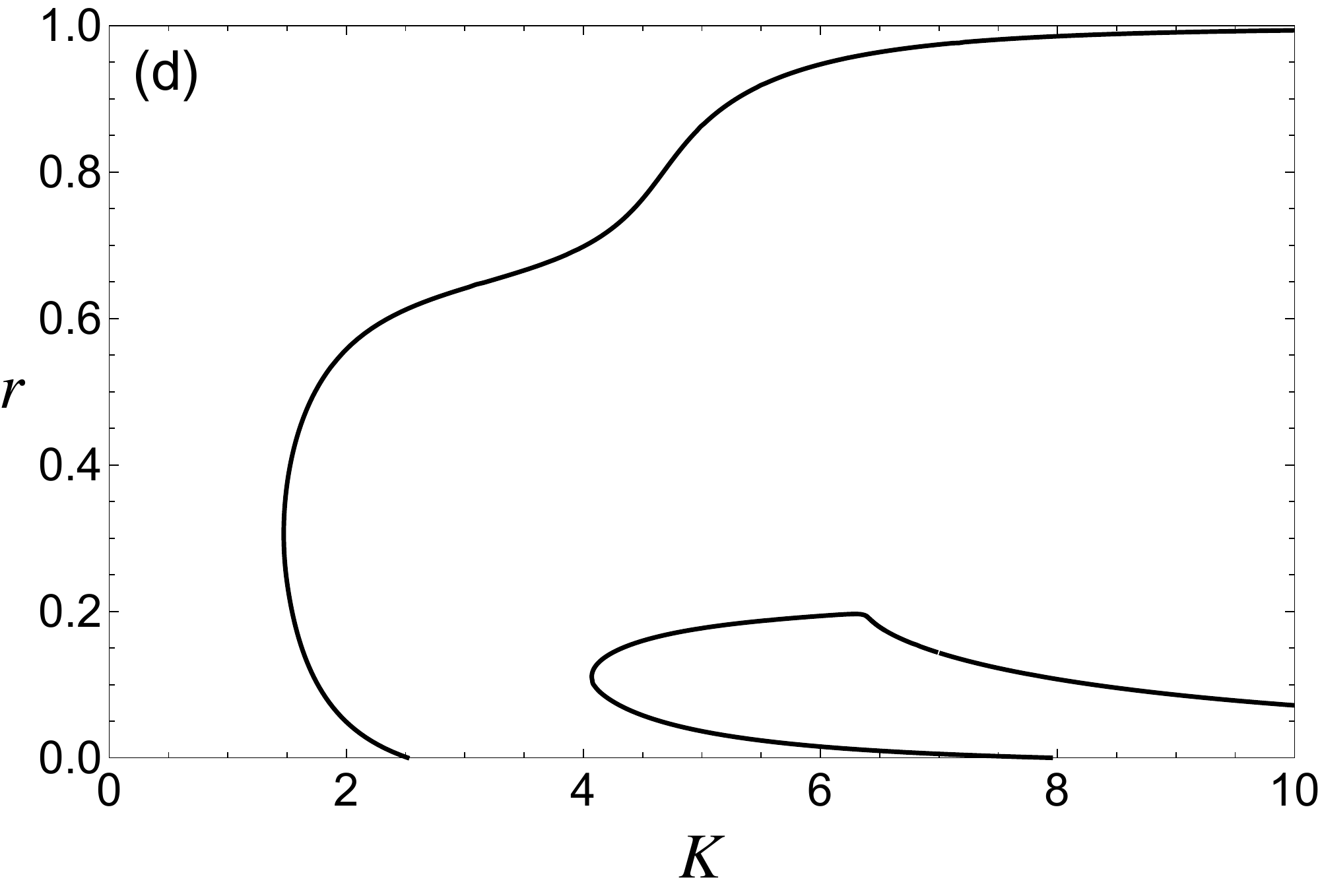}}
  \caption{Steady state solutions of the self-consistent equations for a system with bimodal density $g(\Omega)$, Eq.~\eqref{eq_bimodal_density}, $m=2$, $D=1$, and $b_P=b_S$, corresponding to the backward process [B].
  Panel (a) depicts $g(\Omega)$.
  Panel (b) shows the zero-sets of $F_2(q,\Omega^r)$, Eq.~\eqref{eq_self_consistent_calculation_imag}, with black curves.
  The gray areas represent points $(q,\Omega^r)$ where $F_1(q,\Omega^r) \le 0$ and thus cannot correspond to solutions of Eq.~\eqref{eq_self_consistent_calculation} even if the satisfy $F_2(q,\Omega^r) = 0$.
  We note the existence of three solution branches $\Omega^r(q)$ of Eq.~\eqref{eq_self_consistent_calculation_imag} for small $q$;
  for larger $q$ only one branch remains.
  Panel (c) shows the families $(K(q),r(q),\Omega^r(q))$ obtained through Eq.~\eqref{eq_self_consistent_calculation_real} as described in the text.
  Panel (d) shows the projection of the families from panel (c) to the $(K,r)$-plane.
  }
  \label{fig_self_consistent}
  \label{fig_self_consistent_bimodal_example}
\end{figure*}

As an example of this approach we consider a system with the bimodal density function
\begin{align}\label{eq_bimodal_density}
  g(\Omega) = \frac{3}{10} \sqrt{\frac{2}{\pi}} \exp[-2(\Omega+1)^2]
   + \frac{7}{10} \sqrt{\frac{2}{\pi}} \exp[-2(\Omega-1)^2],
\end{align}
see Fig.~\ref{fig_self_consistent_bimodal_example}(a), and fix the parameter values to $D=1$ and $m = 2$. The solution set of $F_2(q,\Omega^r) = 0$ is shown in Fig.~\ref{fig_self_consistent_bimodal_example}(b) and the corresponding solutions are depicted in Fig.~\ref{fig_self_consistent_bimodal_example}(c) in the $(K,r,\Omega^r)$-space and projected onto the $(K,r)$ plane in Fig.~\ref{fig_self_consistent_bimodal_example}(d). Note the existence of more than one $K$-parameterized families $(r(K),\Omega^r(K))$.
Moreover, note in Fig.~\ref{fig_self_consistent_bimodal_example}(b) that some points $(q,\Omega^r)$ satisfying $F_2(q,\Omega^r) = 0$ lie in the region where $F_1(q,\Omega^r) \le 0$, represented by the gray color in Fig.~\ref{fig_self_consistent_bimodal_example}(b).
Such points cannot represent a solution of the self-consistent equation since they give $1/K \le 0$.
Therefore, they must be rejected and they do not contribute to subsequent panels (c) and (d) in Fig.~\ref{fig_self_consistent_bimodal_example}.

In the rest of this section we explore in more detail the properties of steady states obtained as solutions of the self-consistent equation, Eq.~\eqref{eq_self_consistent_calculation}, for arbitrary distributions $g(\Omega)$. 
In particular, we discuss the existence of multiple solution branches of the self-consistent equations, the bifurcation points from the incoherent state (transition points), and steady states beyond the forward and backward processes.
In the subsequent Sec.~\ref{section_symmetry} we focus the discussion of the steady state solutions and their properties tox the case of unimodal symmetric distributions.

\subsection{Multiple solution branches}

Solutions to the self-consistent equation, Eq.~\eqref{eq_self_consistent_calculation}, for second-order oscillators can have multiple branches.
This feature is a natural consequence of the nonlinear nature of the self-consistent equations.

First, for a given value of $q$, there may be multiple solution branches $\Omega^r(q)$ of the equation $F_2(q,\Omega^r) = 0$, Eq.~\eqref{eq_self_consistent_calculation_imag}.
The number of these branches is always odd (counting multiplicity), as a consequence of the continuity of $F_2(q,\Omega^r)$, and the fact that there is $M > 0$ such that $F_2(q,\Omega^r) < 0$ for $\Omega^r > M$ and $F_2(q,\Omega^r) > 0$ for $\Omega^r < -M$.
Second, for each solution branch $\Omega_r(q)$, we have corresponding one-parameter families of steady states $(K(q),r(q),\Omega^r(q))$ through Eq.~\eqref{eq_self_consistent_calculation_real}.
For each such family we can solve to obtain $r$ as a function of $K$. 
However, for each family there may be more than one such branches $r(K)$.
This is depicted in Fig.~\ref{fig_self_consistent_bimodal_example}(d) for the bimodal distribution, Eq.~\eqref{eq_bimodal_density}, and in Fig.~\ref{fig_self_consistent_symmetry} for a Gaussian distribution, Eq.~\eqref{eq_gauss}, with $\sigma = 1$.

In addition, since $F_1(q,\Omega^r)$ is bounded we conclude that $K$ cannot take values smaller than some $K_\text{min} > 0$ for solutions of the self-consistent equation, Eq.~\eqref{eq_self_consistent_calculation}, or, equivalently, for the non-trivial solutions of Eq.~\eqref{self_consistent_eq_separate}. 
This implies that the only solution that is possible for $K < K_\text{min}$ is the trivial solution $r = 0$ and other branches are the result of bifurcations that occur at values of $K$ larger than $K_\text{min}$.

\begin{remark}
  The first-order Kuramoto model also exhibits multiple branches of steady state solutions (multi-stability) if we consider non-unimodal natural frequency distributions \cite{Acebron2005}, phase shifts \cite{Omel2013}, or complex network topologies \cite{Manik2017cycle}.
\end{remark}

\subsection{Transition points}

The trivial solution $r = 0$ represents the incoherent state.
We are interested at the \emph{transition points}, that is, the values $K_c$ of the coupling strength where non-trivial solutions of the self-consistent equation either merge with or detach from the incoherent state for $r > 0$. 
Such transition points are important since often they coincide with the loss of stability of the incoherent state and the occurrence of a phase transition in the forward process and can also be called forward critical points \cite{Zou2014}.
Note that the transition points $K_c$ do not necessarily correspond to the minimum value of the coupling strength for which the system has non-trivial solutions, as can be seen in the examples in Fig.~\ref{fig_self_consistent_bimodal_example}(d) and in Fig.~\ref{fig_self_consistent_symmetry}.

Using Eq.~\eqref{eq_self_consistent_calculation}, we can determine the transition points taking the limit $q \to 0^+$ corresponding to $r \to 0^+$.
When $q \to 0^+$, we have $a \to \infty$ and hence $b_P(a)=1$ for both forward and backward processes.
This implies that the computed transition points are the same for forward and backward processes and for all the intermediate steady states, see Sec. \ref{sec_dependence}.
However, we must stress that only in the forward process the transition point $K_c$ is the value of $K$ where the incoherent state becomes unstable and the system moves to a stable non-trivial steady state solution.
In the backward process the system may pass from a stable non-trivial steady state solution to the stable incoherent state for values of $K$ smaller than $K_c$.

The transition points $(\Omega_c^r,K_c)$ are determined through non-trivial solutions of the self-consistent equation, Eq.~\eqref{eq_self_consistent_calculation}, which we rewrite as
  \begin{align*}
  \frac{1}{K_c} &=\lim_{q \to 0^+}
  \int_{-1}^{1}
  g(qb+D\Omega^r_c)\sqrt{1-b^2}
  \, db \\
  & \quad -\lim_{q \to 0^+}
  \left[\int_{1}^{\infty}+\int_{-\infty}^{-1}\right]
  g(qb+D\Omega^r_c)
  \frac{a^2}{2(b^2+a^4)}
  \, db,
  \\
  0 &= \lim_{q \to 0^+}
  \int_{-1}^{1}
  g(qb+D\Omega^r_c) 
  b
  \, db \\
  & \quad + \lim_{q \to 0^+}
  \left[\int_{1}^{\infty}+\int_{-\infty}^{-1}\right]
  g(qb+D\Omega^r_c)
  \frac{a^4}{2b(b^2+a^4)}
  \, db. 
  \end{align*}

The change of variables $x=qb$, the introduction of the \emph{reduced mass} $\mu = m/D^2$, and subsequent calculations bring the previous equations to the form
\begin{subequations}
  \label{eq_transition_point}
  \begin{align}
  & \frac{1}{K_c} =\frac{\pi}{2}g(D\Omega^r_c)
  -\intR
  g(x+D\Omega^r_c)
  \frac{\mu}{2 (1+\mu^2 x^2)}
  \, dx,
  \label{eq_transition_point_real}
  \\
  & 0 =\lim_{q \to 0^+}
  \int_{q}^{\infty}
  \frac{g(x+D\Omega^r_c)-g(-x+D\Omega^r_c)}{2x}
  \frac{1}{1+\mu^2x^2}
  \, dx.
  \label{eq_transition_point_imag}
  \end{align}
\end{subequations}
If the steady state branch that bifurcates at $K = K_c$ from the incoherent state is unstable then the transition between the incoherent state and corresponding coherent state is discontinuous. 
Otherwise, the transition is continuous.

\begin{remark}
  Fig.~\ref{fig_self_consistent_bimodal_example}(b) shows that it is possible that $\lim_{q \to 0^+} F_1(q, \Omega^r(q)) < 0$ and thus $K_c < 0$.
  We reject such solutions since for $q > 0$ (as we consider here) they give the non-physical $r < 0$.
  Consider the situation depicted in Fig.~\ref{fig_self_consistent_bimodal_example}(b) where a curve $\mathcal C$ of solutions of $F_2(q,\Omega^r) = 0$ enters the region $F_1(q, \Omega^r) \le 0$ by crossing the zero-set of $F_1(q, \Omega^r)$ (dashed curve in Fig.~\ref{fig_self_consistent_bimodal_example}(b)) at a point $(q_0, \Omega^r_0)$.
  Then, clearly, only the part $\mathcal C_+$ of $\mathcal C $ where $F_1(q, \Omega^r) > 0$ can be considered.
  Consider a point $(q,\Omega^r)$ on $\mathcal C_+$ that approaches $(q_0,\Omega^r_0)$.
  Then the value of $F_1(q,\Omega^r)$ approaches $0$ (while positive), and thus $K = [F_1(q,\Omega^r)]^{-1}$ approaches $\infty$.
  This implies that in the $(K,r)$ plane we obtain a family $(K(q),r(q))$ with $K(q) \to \infty$ and $r(q) \to 0$ as $q \to q_0$ in such a way so that $K(q) r(q) \to q_0$ as $q \to q_0$.
  In other words, for large enough $K$ the corresponding curve $r(K)$ becomes asymptotic to the hyperbola $K r = q_0$.
\end{remark}

\begin{remark}
  Compared with the Kuramoto model where $K_c=2/(\pi g(D\Omega^r_c))$, the effect of inertias in Eq.~\eqref{eq_transition_point_real} is always to decrease the value of $1/K_c$ since the integral in this equation is non-negative.
  Hence with the same $\Omega_c^r$, the critical coupling strength $K_c$ for second-order oscillators is always larger than the one for Kuramoto oscillators.
\end{remark}

\subsection{Steady states beyond the forward and backward processes}
\label{sec_dependence}

For second-order oscillators, a crucial complication is the existence of the bistable state and the corresponding complicated basins of attraction.
Restricting our attention to the forward and backward processes, leading to Eq.~\eqref{self_consistent_eq_separate}, this complication is avoided since then the locked and running oscillators are separated by the boundaries of the bistable region.

The steady states attained in the forward and backward processes is a special collection of steady states.
In general, for other processes with arbitrary choice of initial states it is hard to analytically find the boundary between locked and running oscillators and consequently obtain the corresponding steady states. 
However, with different initial states, the oscillators will always separate into two groups.
The corresponding fractions can be defined as $C_l(b;a)$ and $C_r(b;a)$ for locked and running groups respectively, with normalization condition $C_l(b;a)+C_r(b;a)=1$.
In the special case of the forward process we have $C_l(b;a) = \mathbf{1}_{l,b_S}$ where $\mathbf{1}_{l,b_P}$ takes the value $1$ if $|b| < b_P(a)$ and $0$ otherwise.
In the backward process we similarly have $C_l(b;a) = \mathbf{1}_{l,b_L}$.
In terms of the fractions $C_l(b;a)$ and $C_r(b;a)$ the self-consistent equation reads
\begin{subequations}
  \label{eq_self_consistent_general}
  \begin{align}
  r &= \intR
  G(b) 
  \left[C_l(b;a)\sqrt{1-b^2}-C_r(b;a)\frac{a^2}{2(b^2+a^4)}\right]
  db,
  \label{eq_self_consistent_general_real}
  \\
  0 &= \intR
  G(b)
  \left[C_l(b;a)b+C_r(b;a)\frac{a^4}{2b(b^2+a^4)}\right]
  db.
  \label{eq_self_consistent_general_imag}
  \end{align}
\end{subequations}
Even though we cannot easily determine $C_l(b;a)$, we note that
\begin{align*}
  \mathbf{1}_{r,b_S} \le C_l(b;a) \le \mathbf{1}_{r,b_L}.
\end{align*}
Therefore, different possibilities can be viewed as intermediate between the two considered processes.
In particular, we can consider a boundary function $b_P(a)$ given as convex combination of the boundaries for the two processes, that is,
\begin{align*}
  b_P(a) = c \, b_S(a) + (1-c) \, b_L(a).
\end{align*}

The previous discussion implies that, for fixed parameters $(m, D, K)$ and fixed frequency distribution $g(\Omega)$, different initial states may reach different steady states.
This is further discussed in Sec.~\ref{sec_margin_region} and demonstrated in Fig.~\ref{fig_dependence} for a symmetric unimodal distribution.

\subsection{Frequency scaling and scaled inertia}
\label{sec_scale_free_inertia}

Consider a frequency distribution $g_s(\Omega)$ that depends on a \emph{scale parameter} $s > 0$ so that
\begin{align*}
  g_s(\Omega) = \frac{1}{s} g_1\left(\frac{\Omega}{s}\right).
\end{align*}
Typical examples are the Gaussian distribution, Eq.~\eqref{eq_gauss}, where $s = \sigma$, and the Lorentz distribution, Eq.~\eqref{eq_lorentz}, where $s = \gamma$.

Suppose that for inertia $\mu_1$ and distribution $g_1$ one finds a steady state solution of the self-consistent equation (not necessarily one obtained through a forward or backward process), characterized by the parameters $(q_1, w_1, K_1, r_1)$.
Here we introduce the parameter $w = D \Omega^r$ since $\Omega^r$ appears in the self-consistent equation only through $w$. Then a straightforward computation shows that for given inertia $\mu_s = s^{-1} \mu_1$ and for given distribution $g_s$ there is a steady state solution characterized by parameters $(q_s, w_s, K_s, r_s)$ with
\begin{align*}
  q_s = s q_1, \quad w_s = s w_1, \quad K_s = s K_1, \quad r_s = r_1.
\end{align*}
This property of steady-state solutions allows the translation of results from $s=1$ to any value of $s > 0$.
In particular, this allows the straightforward translation of the numerical results in Sec.~\ref{section_symmetry}, which have been obtained for a Gaussian distribution with $\sigma = 1$, to the case of arbitrary $\sigma > 0$.

Moreover, this observation suggests that we should introduce a more natural notion of inertia,  the \emph{scaled inertia}
\begin{align*}
  \nu = s \mu, 
\end{align*}
so that $\nu_s = s \mu_s = \mu_1 = \nu_1$ is invariant under the scaling transformation.
In what follows we do not directly use $\nu$ since we are interested in distributions that do not necessarily depend on a scale parameter.

\section{Symmetric unimodal natural frequency distribution}
\label{section_symmetry}

In this section, we consider the system with symmetric unimodal density function $g(\Omega)$.
Note that for a single oscillator with natural frequency $\Omega$, we can describe its dynamics in a frame rotating with frequency $\Omega'$ as having a new natural frequency $\Omega - D \Omega'$.
Because of this property, we can assume that the median (and, when defined, also the mean) of $g(\Omega)$ is zero.
Moreover, we have $g(\Omega) = g(-\Omega)$, and $g(\Omega_1) \le g(\Omega_2)$ if $\Omega_1 \ge \Omega_2 \ge 0$ from the unimodal property.
Two typical symmetric unimodal distributions are the Gaussian distribution
\begin{align}\label{eq_gauss}
  g(\Omega) = \frac{1}{\sqrt{2 \pi \sigma^2}} \exp\Big( -\frac{\Omega^2}{2\sigma^2} \Big),
\end{align}
and the Lorentz (or Cauchy) distribution
\begin{align}\label{eq_lorentz}
  g(\Omega) = \frac{1}{\pi} \frac{\gamma}{\Omega^2 + \gamma^2}.
\end{align}

\begin{figure*}[tb]
  \centering
  {\includegraphics[width=0.48\textwidth]{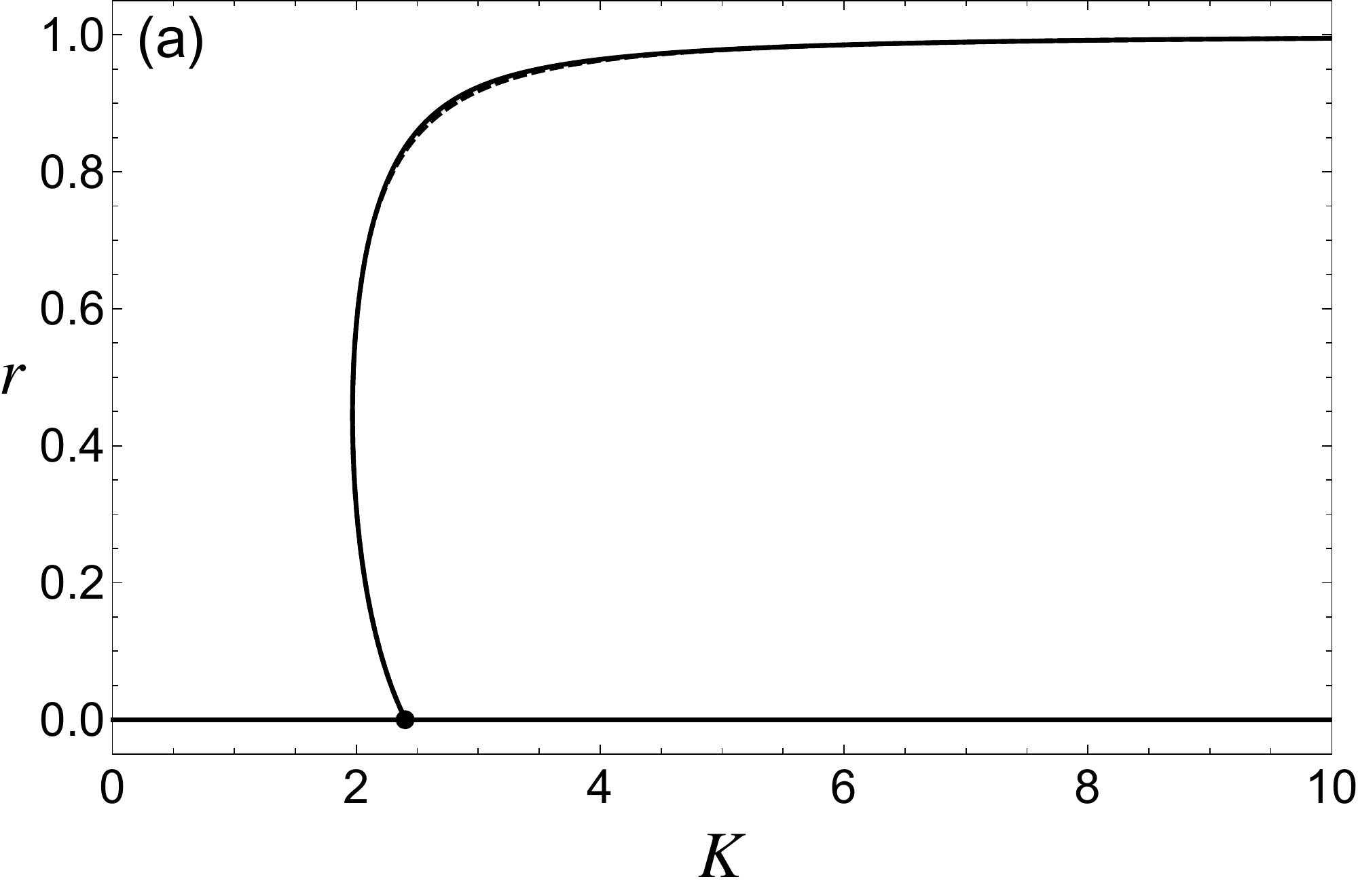}}
  \hfill
  {\includegraphics[width=0.48\textwidth]{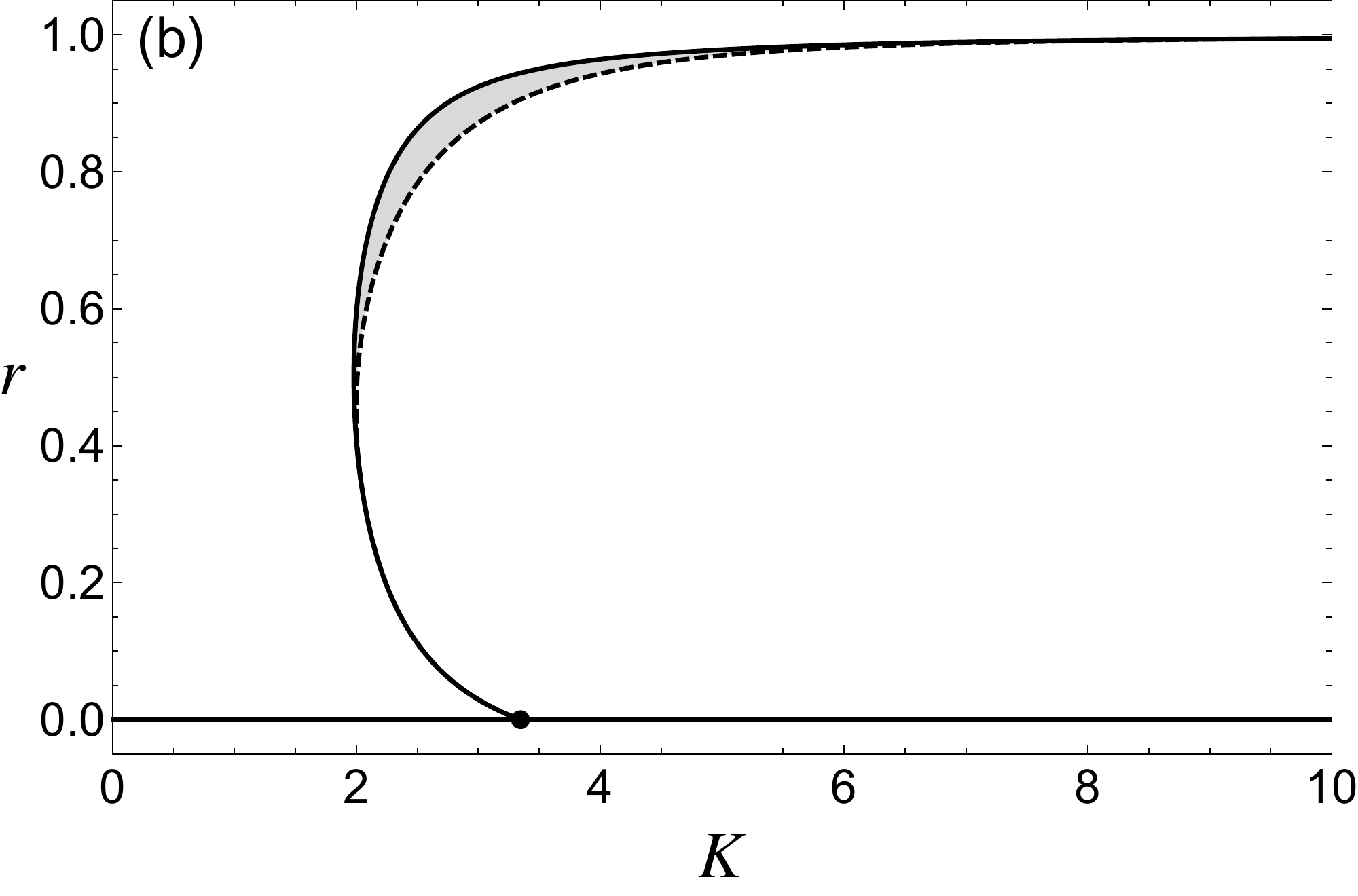}}
  \\
  {\includegraphics[width=0.48\textwidth]{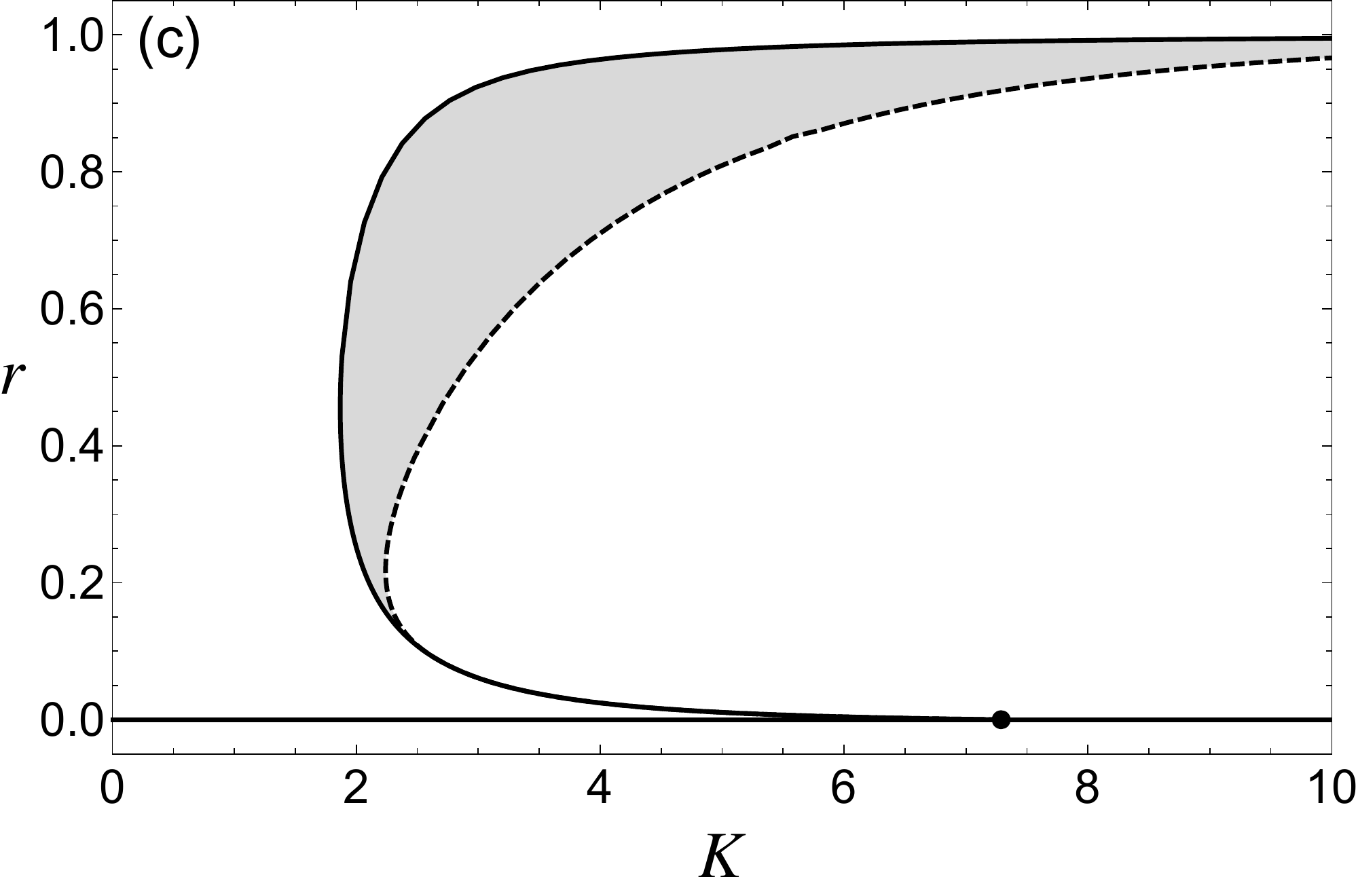}}
  \hfill
  {\includegraphics[width=0.48\textwidth]{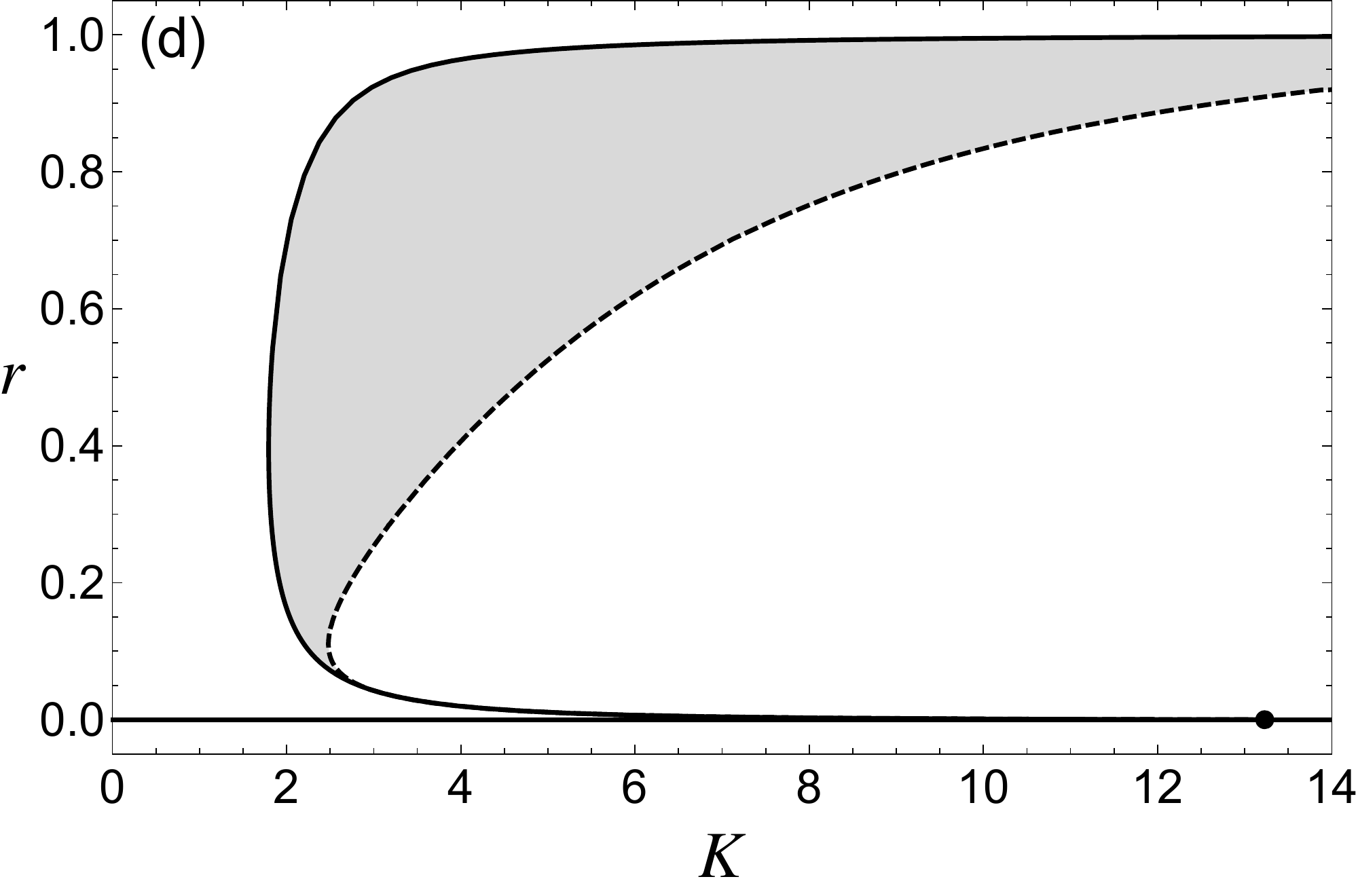}}
  \caption{The steady state solutions obtained by numerical integration of Eq.~\eqref{eq_self_consistent_symmetry} for oscillators with Gaussian $g(\Omega)$, $\sigma = 1$,
  depicted together with the incoherent state $r=0$.
  The damping coefficient is $D=1$.
  From top left to bottom right: (a) $m = 0.5$; (b) $m=1$; (c) $m=3$; (d) $m=6$.
  Solid lines (except $r=0$) represent the backward process with $b_P=b_L$, and dashed lines the forward process with $b_P=b_S$.
  The gray-shaded area represents the margin region $\mathcal{M}$ discussed in Sec.~\ref{sec_margin_region}.
  Transition points where steady state solutions merge with or detach from the incoherent state $r=0$ are marked in the picture.
  }
  \label{fig_self_consistent_symmetry}
\end{figure*}

\begin{figure*}[tb]
  \centering
  {\includegraphics[width=0.48\textwidth]{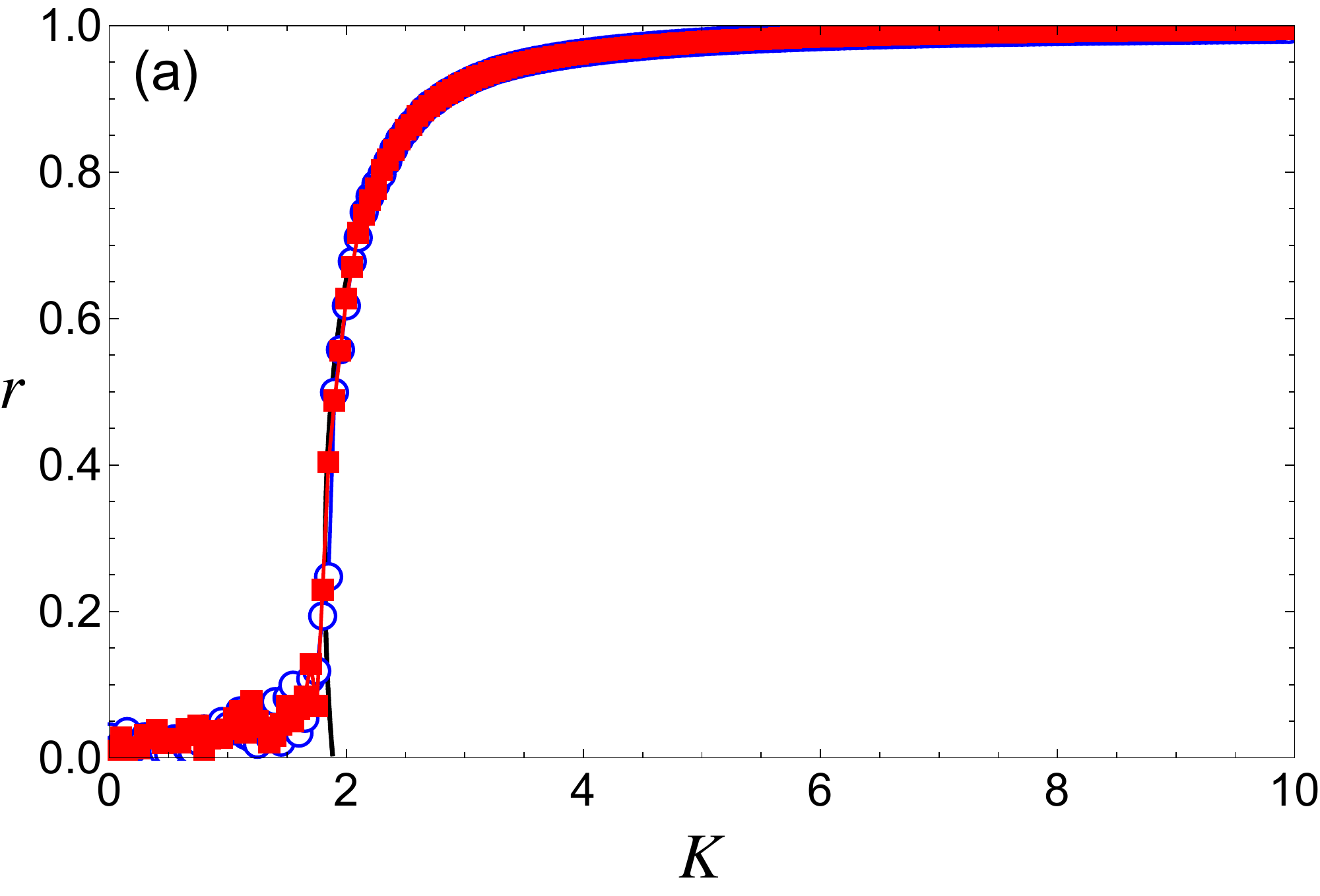}}
  \hfill
  {\includegraphics[width=0.48\textwidth]{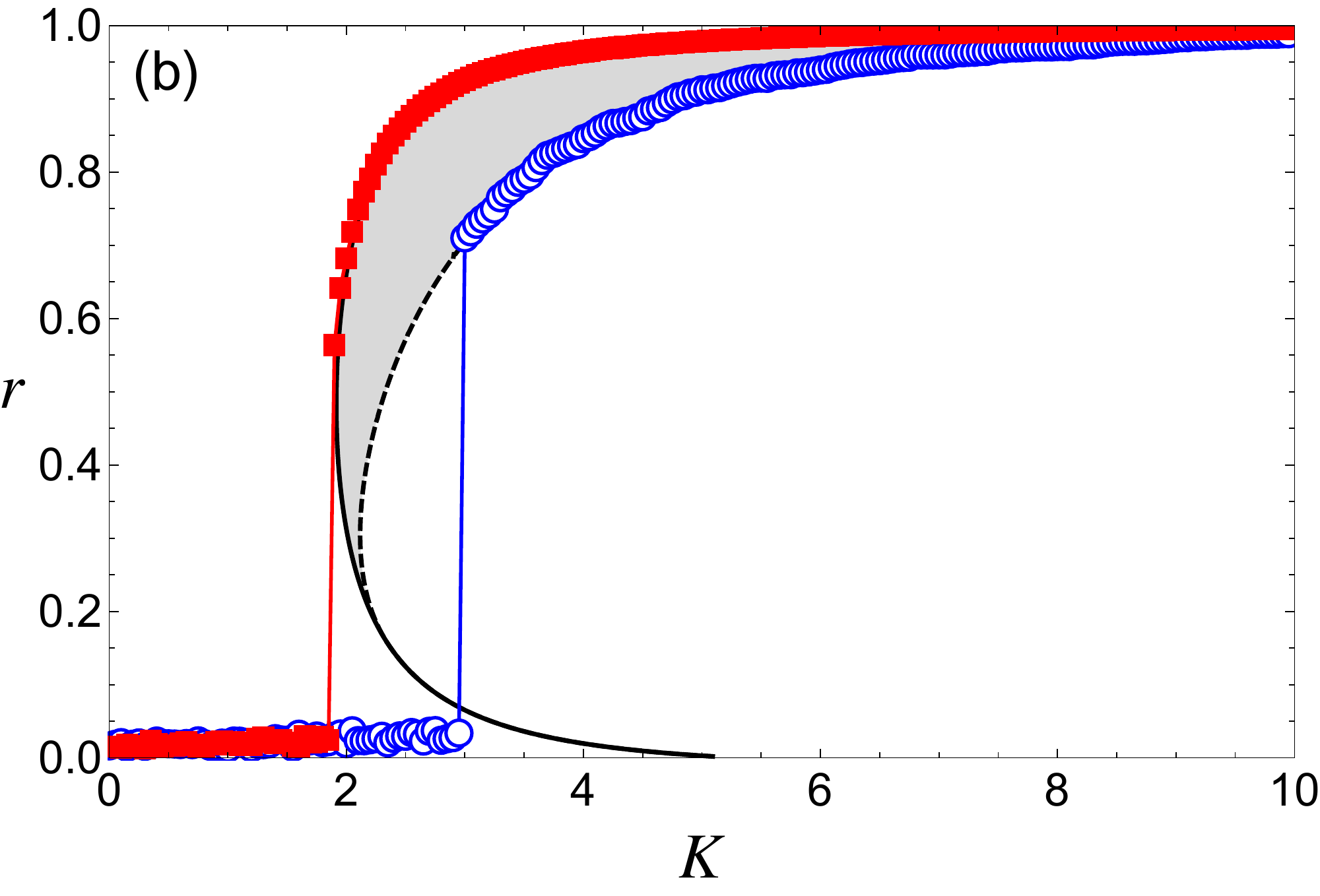}}
  \caption{Numerical simulations of the backward and forward processes for $N=5000$ oscillators with (a) $m=0.2$ and (b) $m=2$ where $D=1$, and $g(\Omega)$ is the Gaussian distribution, Eq.~\eqref{eq_gauss}, with $\sigma = 1$.
  Solid lines represent the backward process with $b_P=b_L$, and dashed lines the forward process with $b_P=b_S$.
  The numerical results for the backward processes are represented by (red filled) squares and for the forward processes by (blue) circles.
  In panel (a) the points for the two processes largely overlap.}
  \label{fig_numerical_one}
\end{figure*}

\subsection{Self-consistent equations}

One crucial characteristic of the coupled oscillators with symmetric unimodal $g(\Omega)$ is that there is only one solution $\Omega^r(q) = 0$ of the equation $F_2(q,\Omega^r)=0$.
To show this, we rewrite Eq.~\eqref{eq_self_consistent_calculation_imag} as
\begin{align}\label{eq_rewrite_sc_imag}
\begin{aligned}
0
& = F_2(q,\Omega^r) \\
& = \int_{0}^{\infty} \big[ g(qb+D\Omega^r)-g(-qb+D\Omega^r) \big] \, b W(|b|) \, db,
\end{aligned}
\end{align}
where $W(|b|)$ is the even positive function given by 
\begin{align*}
  W(|b|)
  =
  \mathbf{1}_l + \mathbf{1}_r \frac{a^4}{2b^2(b^2+a^4)}
  > 0.
\end{align*}
Since the distribution $g(\Omega)$ is symmetric and unimodal, we have for $qb \ne 0$ that $g(qb+D\Omega^r)-g(-qb+D\Omega^r)=0$ if and only if $\Omega^r = 0$.
Hence the only solution of Eq.~\eqref{eq_rewrite_sc_imag}, or equivalently Eq.~\eqref{eq_self_consistent_calculation_imag}, is $\Omega^r=0$.

\begin{remark}
  Note that for a symmetric (not necessarily unimodal) $g(\Omega)$ the function $F_1(q,\Omega^r)$ is even in $\Omega^r$ while $F_2(q,\Omega^r)$ is odd in $\Omega^r$.
  The latter property implies that $F_2(q,0) = 0$ for all $q$, while the former property then implies that the corresponding value $F_1(q,0)$ is a local maximum or minimum value of $F_1(q,\Omega^r)$ for fixed $q$. 
\end{remark}

With the substitution $\Omega^r = 0$, Eq.~\eqref{eq_self_consistent_calculation_real} reads 
\begin{equation}
\frac{1}{K}
= F_1(q, 0)
=\intR  g(qb) 
\left[
\mathbf{1}_l \sqrt{1-b^2}
- \mathbf{1}_r \frac{a^2}{2(b^2+a^4)}
\right]
db,
\label{eq_self_consistent_symmetry}
\end{equation}
where we remind that $b_P = b_S$ for the forward process and $b_P = b_L = 1$ for the backward process.
The solutions to the self-consistent equation, Eq.~\eqref{eq_self_consistent_symmetry}, for different values of $\mu = m / D^2$ and a Gaussian distribution with $\sigma = 1$ are shown in Fig.~\ref{fig_self_consistent_symmetry}.

Even though for symmetric unimodal distributions $g(\Omega)$ the self-consistent equation $F_2(q,\Omega^r)=0$ has only one solution $\Omega^r(q)=0$, we can still obtain multiple solution branches $r(K)$ from the self-consistent equation $F_1(Kr,0)=1/K$, Eq.~\eqref{eq_self_consistent_symmetry}.
Multiplying both sides of the last equation by $q = Kr$ and using the original parameters (alternatively,  substituting the original parameters in Eq.~\eqref{eq_self_consistent_separate_real} and then setting $\Omega^r = 0$) we obtain the equation
\begin{subequations}
\begin{align}
  r = F(r, K),  
  \label{eq_self_consistent_symmetry2}
\end{align}
where $F(r,K)$ is given by
\begin{align}
\begin{aligned}
  F(r,K)
  & = 
  \int_{|\Omega| < Kr\,b_P(1/\sqrt{Kr\mu})} 
  g(\Omega) \sqrt{1-\frac{\Omega^2}{K^2r^2}} \, d\Omega \\
  & - \int_{|\Omega| > Kr\,b_P(1/\sqrt{Kr\mu})}
  g(\Omega) \frac{Kr\mu}{2(1 + \mu^2\Omega^2)} \, d\Omega.
\end{aligned}
\end{align}
\end{subequations}
We note that Eq.~\eqref{eq_self_consistent_symmetry2} has the trivial solution $r=0$, that is, $F(0,K) = 0$.
Moreover, we have
\begin{align*}
F(1,K) 
& < \int_{|\Omega|< K\,b_P(1/\sqrt{K\mu})}
g(\Omega) 
\sqrt{1-\frac{\Omega^2}{K^2}} \, d\Omega \\
& < \int_{|\Omega|< K\,b_P(1/\sqrt{K\mu})}
g(\Omega) \, d\Omega
< 1.
\end{align*}
Hence there is at least one solution of Eq.~\eqref{eq_self_consistent_symmetry2} with $0 \le r \le 1$. 

To check the analysis of steady states, we have performed several numerical simulations. 
We have numerically calculated the dynamics of a network with $N=5000$ oscillators, following Eq.~\eqref{Dynamic-of-single}, using the fourth order Runge-Kutta method with fixed-size integration time-step $dt = 10^{-3}$.
The natural frequency $\Omega_i$ for each oscillator is chosen randomly from a Gaussian distribution $g(\Omega)$ with $\sigma=1$.
At a given coupling strength $K$, after a transient period $t_0 = 40$, we calculate the order parameter $r$ according to the definition in Eq.~\eqref{order_definition_eq} as the average value over a measurement period $\Delta t=4$.
In the forward and backward processes, we take $dK = 10^{-2}$ and $dK = -10^{-2}$ respectively as the increasing and decreasing coupling strength steps.
In each step, the initial states of all the oscillators are the last states in the previous step.
In the backward and forward processes, the initial states of the first step are chosen randomly from $\theta(0) \in[0,2\pi]$, $\dot{\theta}(0)\in[0,1]$ and $\theta(0)\in [0,0.02\pi]$, $\dot{\theta}(0)\in[0,1]$ respectively.

The phase transitions in the forward and backward processes are shown in Fig.~\ref{fig_numerical_one}.
for oscillators with inertias $m=0.2$ in panel (a) and $m=2$ in panel (b).
In all simulations $D = 1$. 
The figures show that our analytical results coincide with the numerical ones quite well and much better than the analytical results given in \cite{Tanaka1997,Tanaka1997a}.
The error in the location of the transition point is due to the finite number ($N=5000$) of oscillators used in the numerical simulations, whereas the self-consistent analysis is based on the limit $N \to \infty$; see~\cite{Olmi2014} for a more detailed discussion of this phenomenon.

\subsection{Transition points}

Near the incoherent state, that is for $r \approxeq 0$, we have $b_S = b_L = 1$, and there is no bistable behavior.
Substituting $\Omega^r=0$ into Eq.~\eqref{eq_transition_point_real} we obtain the critical value $K_c(\mu)$ of $K$ as a function of the reduced inertia $\mu = m / D^2$.
This is given by
\begin{equation}
\label{eq_K_C}
  K_c(\mu) = \frac{2}{\displaystyle \pi g(0) - A(\mu)},
\end{equation}
where
\begin{equation}
  A(\mu)
  = \intR \frac{\mu}{1+\mu^2x^2} g(x)\,dx
  = \intR \frac{1}{1+y^2} g \Big(\frac{y}{\mu}\Big) \, dy.
\end{equation}
This critical coupling strength $K_c(\mu)$ coincides with the value of coupling strength where the incoherent state becomes unstable, see \cite{Barre2016}.

Since $g$ is unimodal we have that
\begin{equation}
  \frac{dA}{d\mu}
  = \intR\frac{1}{1+y^2} \Big[-g'\Big(\frac{y}{\mu}\Big) \frac{y}{\mu^2}\Big]\,dy > 0,
\end{equation}
and moreover, $\lim_{\mu \to 0} A(\mu)=0$, and $\lim_{\mu \to \infty} A(\mu) = \pi g(0)$.
Hence we have $0 < A(\mu) <\pi g(0)$ and
\begin{align*}
  K_c > \frac{2}{\pi g(0)}
\end{align*}
for $\mu>0$.
In particular, $K_c$ increases with $\mu$, see Fig.~\ref{fig_self_consistent_symmetry}.
In the limit $\mu \to 0$ (corresponding either to small inertia or to large damping coefficient), one obtains the critical coupling strength of Kuramoto oscillators, $K_c(0) = 2 / (\pi g(0))$.

For Gaussian and for Lorentz distributions the value $K_c(\mu)$ can be explicitly computed.
For the Gaussian distribution, Eq.~\eqref{eq_gauss}, we find
\begin{align*}
  K_c(\mu) = \frac{2 \sqrt{2}}{\sqrt{\pi}} \frac{\sigma}{1 - \exp\left(\frac{1}{2\mu^2\sigma^2}\right) \left(1-\mathop{\mathrm{erf}}\left(\frac{1}{\sqrt{2}\mu\sigma}\right)\right)},
\end{align*}
which for small $\mu > 0$ gives
\begin{align*}
  K_c(\mu) = \frac{2 \sqrt{2}}{\sqrt{\pi}} \sigma + \frac{4}{\pi} \sigma^2 \mu + O(\mu^2),
\end{align*}
while for large $\mu$ it gives
\begin{align*}
  K_c(\mu) = 2 \sigma^2 \mu + \sqrt{\frac{\pi}{2}} \sigma + O\left(\mu^{-1}\right).
\end{align*}
For the Lorentz distribution, Eq.~\eqref{eq_lorentz}, we find
\begin{align*}
  K_c(\mu) = 2 \gamma + 2 \gamma^2 \mu.
\end{align*}

\subsection{Margin region}
\label{sec_margin_region}

\begin{figure}[t]
  \centering
  \includegraphics[width=0.48\textwidth]{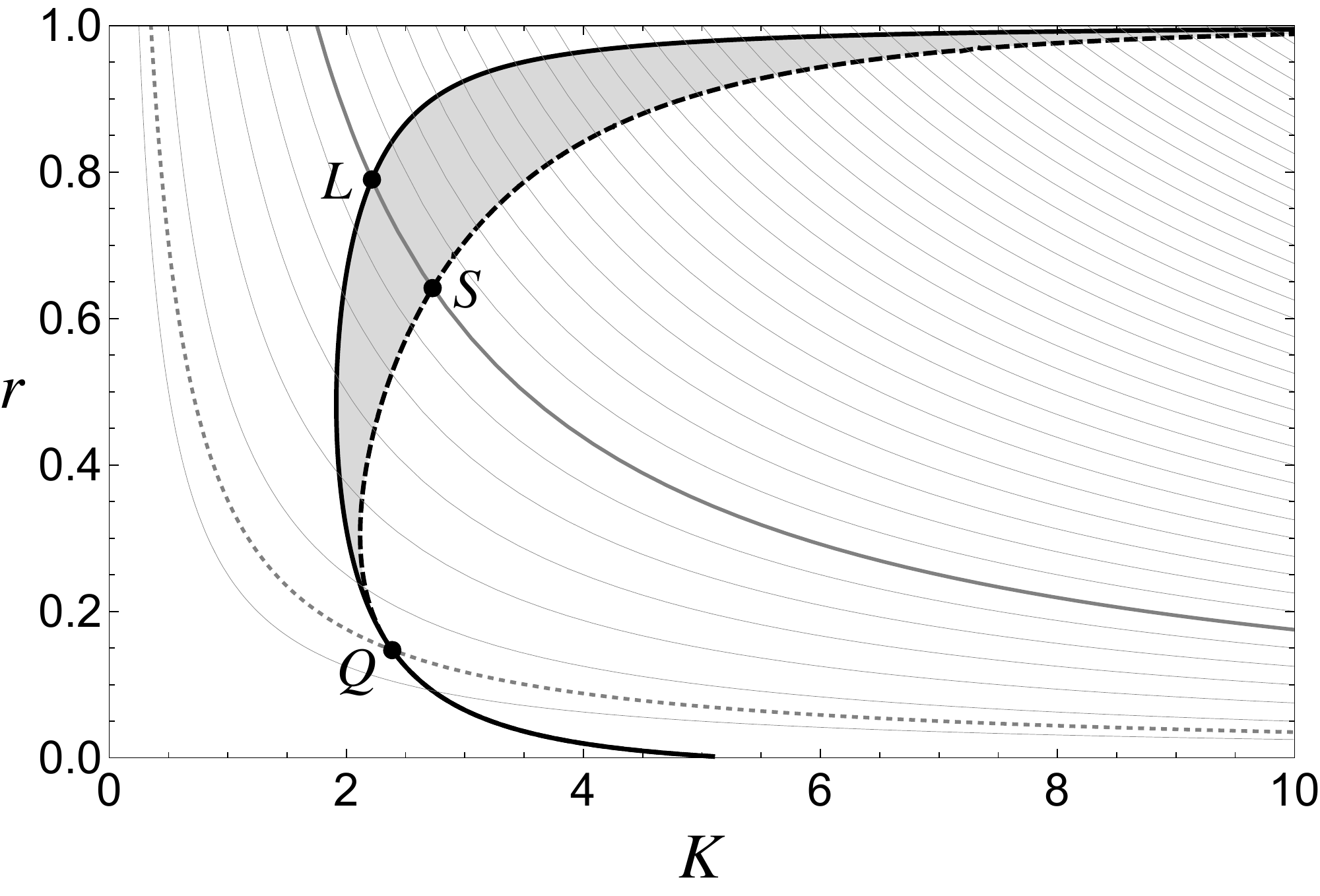}
  \caption{Representation of the margin region, cf.~Fig.~\ref{fig_self_consistent_symmetry}.
  In this picture $m=2$, $D=1$, and $g(\Omega)$ is Gaussian with $\sigma=1$.
  The family of light gray curves represent sets of constant $q = Kr$.
  The margin region appears for $q > q_* \approxeq 0.35$ and is represented by the gray area.
  The set $q = q_*$ is represented by the dotted curve passing through the transitional point $Q$ where the curves for the backward and forward process start differentiating.
  The two steady states marked by $L$ and $S$ correspond to the parameters $(K_L,r_L)$ and $(K_S,r_S)$ respectively, described in Sec.~\ref{sec_margin_region}, for a value of $q$ (here, $q=1.75$).
  }
  \label{fig_margin_region_schematic}
\end{figure}

Recall that for symmetric unimodal distributions $g(\Omega)$ we have $F_2(q,0) = 0$ for all $q > 0$ and any boundary function $b_P(a)$.
This implies that steady states are parameterized by $q \ge 0$ through the relations $K(q) = F_1(q,0)^{-1}$, $r(q) = q F_1(q,0)$. 
Fix a value $q \ge 0$ and let $(K_L(q),r_L(q))$ and $(K_S(q),r_S(q))$ satisfy the self-consistent equations with $b_P = b_L$ and $b_P = b_S$ respectively, assuming that $r_L(q) K_L(q) = r_S(q) K_S(q) = q$, see Fig.~\ref{fig_margin_region_schematic}.
Then we find
\begin{equation}
  \frac{1}{K_L(q)} = F_1(q,0;b_L) \ge F_1(q,0;b_S)=\frac{1}{K_S(q)},
\end{equation}
and subsequently $r_L(q) \ge r_S(q)$.

\begin{figure}[t]
  \centering
  \includegraphics[width=0.48\textwidth]{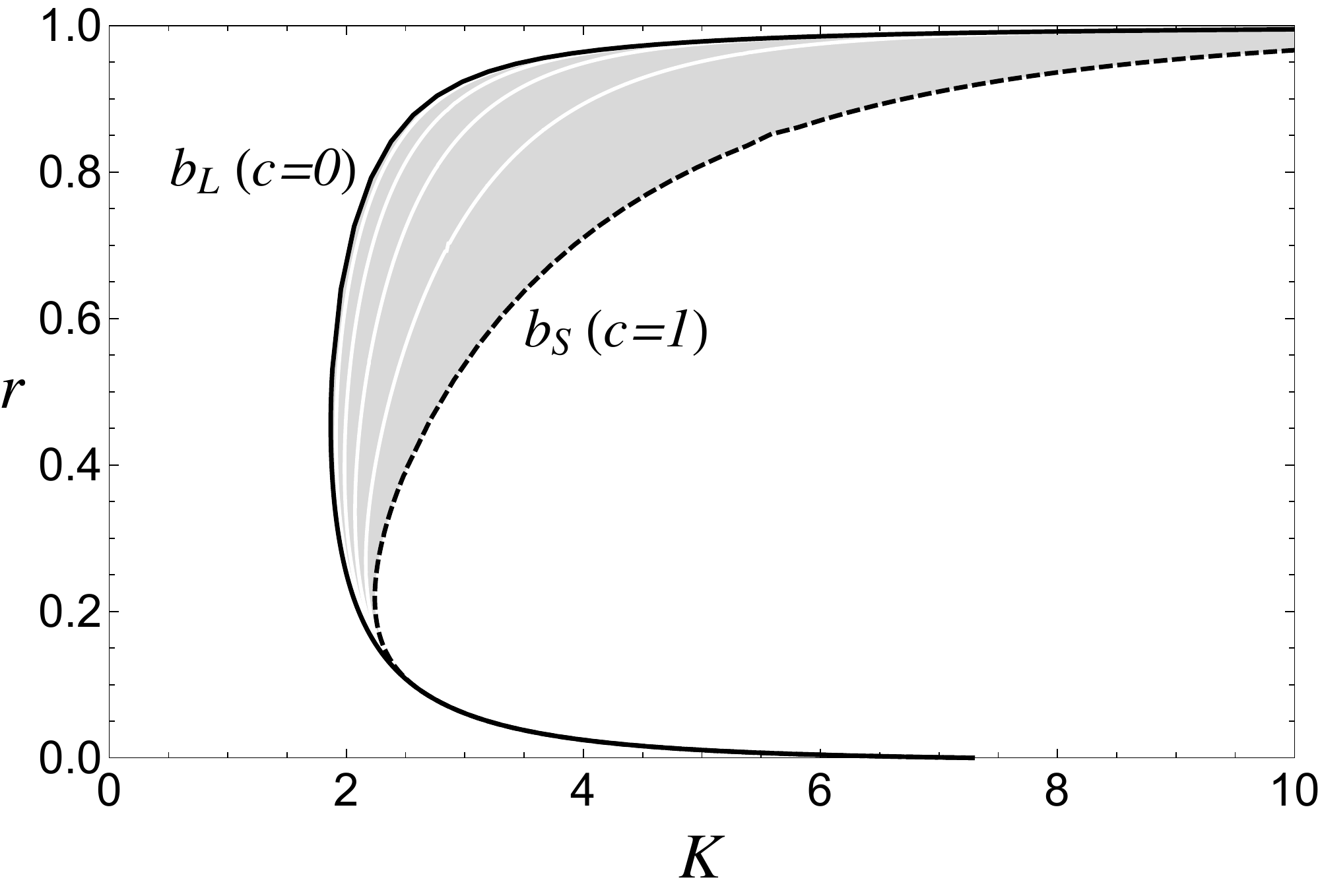}
  \caption{
  The boundaries of the margin region correspond to steady state solutions of the self-consistent equation with $b_P = b_L$ and $b_P = b_S$.
  Points in the margin region can be realized as steady state solutions with $b_P = c\,b_S+(1-c)\,b_L$, $0 < c < 1$.
  In this picture the corresponding curves for $c\in\{0.2,\,0.4,\,0.6,\,0.8\}$ are represented by white curves inside the margin region.
  The parameters are $m=3$, $D=1$, and $g(\Omega)$ is Gaussian with $\sigma=1$.
  }\label{fig_self_consistent_symmetry_2}
\end{figure}

For these two points to be distinct it is necessary that $b_S(a) < b_L(a) = 1$, implying that $a < a_*(1) \approxeq 1.193$ or, equivalently, that 
\begin{equation}
  q > q_* \equiv \frac{1}{a_*(1)^2 \mu} \approxeq \frac{1}{\sqrt{2}\mu},
\label{eq_initial_condition}
\end{equation}
see Fig.~\ref{fig_margin_region_schematic}.
Therefore, we can consider the set $\mathcal{M}$ of steady states characterized by $(r, K)$ with $K \in (K_L(q), K_S(q))$ and $r = q/K$ for $q > q_*$.
We call $\mathcal{M}$ the \emph{margin region}.
Steady states in the margin region can be realized as solutions of the self-consistent equation by considering a boundary function $b_P(a)$ with $b_S(a) < b_P(a) < b_L(a)$, cf.~Sec.~\ref{sec_dependence}, that is, by considering steady states that are not obtained through the forward or backward process, see Fig.~\ref{fig_self_consistent_symmetry_2}.
Thus, with different choices of initial states, the system may attain a steady state in the margin region different than those attained at the forward and backward processes.
This is one crucial difference of second-order oscillators compared to first-order, globally coupled Kuramoto oscillators with unimodular natural frequency distribution.
When one only considers the forward and backward processes, this feature results in the well-known hysteresis of second-order oscillators, see \cite{Tanaka1997a,Tanaka1997} and Fig.~\ref{fig_numerical_one}(b).

\begin{figure}[t]
  \centering
  {\includegraphics[width=0.48\textwidth]{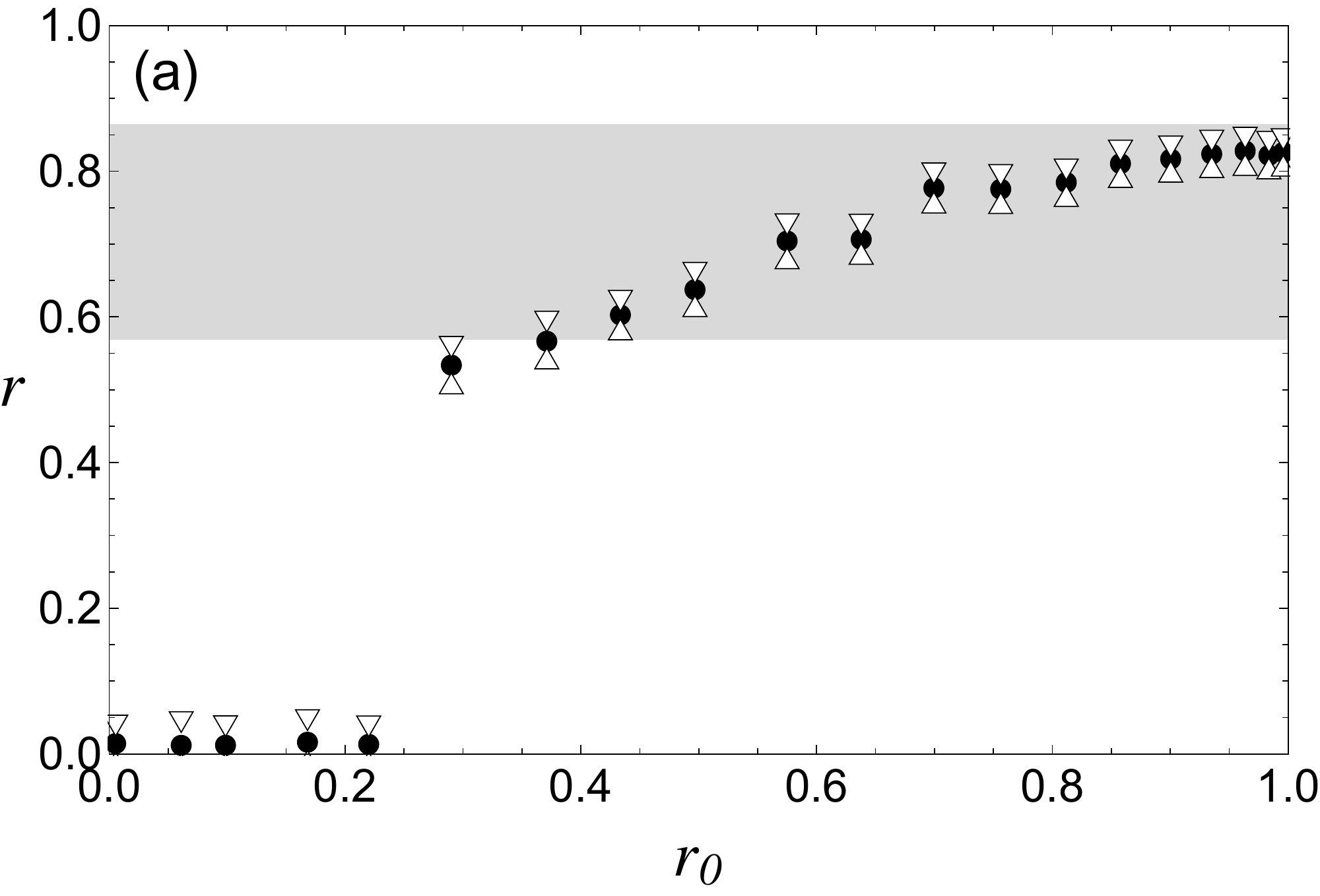}}
  \hfill
  {\includegraphics[width=0.48\textwidth]{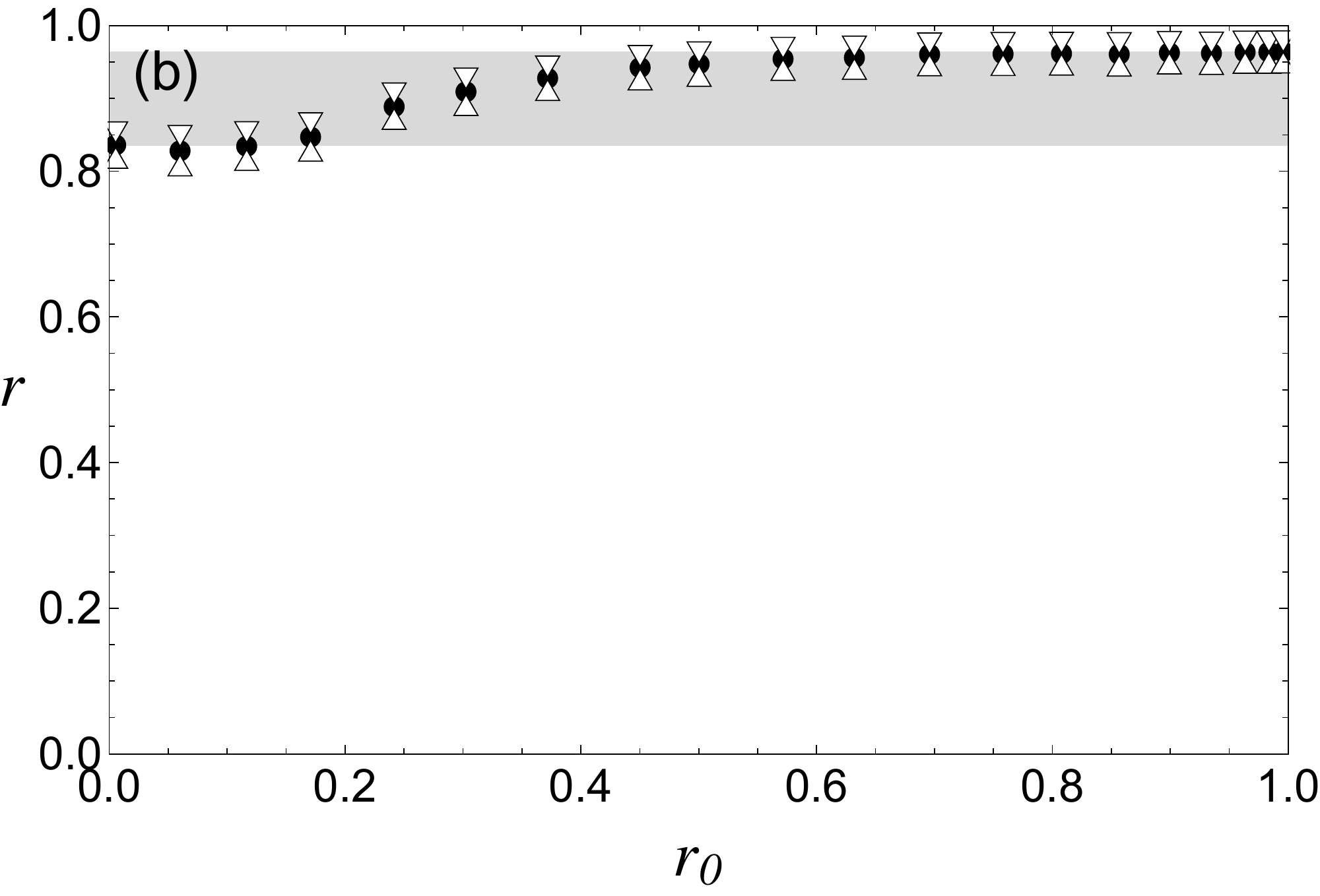}}
  \caption{The initial states dependence of steady states and margin region are shown with initial order parameter $r_0$ and order parameter $r$ (mean: $\bullet$; maximum: $\triangledown$; minimum: $\vartriangle$) with (a) $K=2.5$ and (b) $K=4$. $N=10000$ oscillators are used with $m=2$ and $D=1$.
  The boundaries of the margin region are calculated by setting $b_P=b_L$ and $b_P=b_S$ in Eq.~\eqref{eq_self_consistent_symmetry}.}
  \label{fig_numerical_margin}
  \label{fig_dependence}
\end{figure}

In Fig.~\ref{fig_numerical_margin}, the initial states dependence of the steady states and the corresponding margin regions are shown for oscillators with Gaussian $g(\Omega)$ with $\sigma=1$ for two given coupling strengths $K=2.5$ (a) and $K=4$ (b).
The dynamics of $N=10000$ oscillators with $m=2$, and $D=1$ has been calculated.
The initial phases have been chosen randomly and uniformly in a connected subset of $[0,2\pi]$ so that the corresponding initial order parameter is $r_0$.
The initial phase velocities have been chosen randomly and uniformly in $[-0.5,0.5]$.
After a transient time period $t_0 = 200$, the order parameter $r$ is measured as the time average over a period $\Delta t = 10$.
The maximum and minimum of $r$ is also recorded to show the variation of $r$.
The boundary of regions of stable steady states is calculated with $b_P = b_L$ and $b_P = b_S$ in Eq.~\eqref{eq_self_consistent_symmetry}.
We observe in Fig.~\ref{fig_dependence} that states with different order parameters $r_0$ reach steady states with different $r$ which either correspond the incoherent state or can be found inside the margin region.

\section{Conclusions}\label{section_conclusion}

In this paper, we have considered the self-consistent method for second-order oscillators.
Based on our analysis, and on the obtained self-consistent equations, we have discussed several properties of steady states.
There are several important and novel points in this analysis. 

First, instead of using the original parameters $(m,D,K,r,\Omega^r)$ we have introduced the rescaled parameters $a$ and $b$ in Eq.~\eqref{single_mean_eq}, thus simplifying the analysis of single oscillators but also of the network.

Second, we have given a significantly improved estimate of the limit cycle of the second-order oscillators, where the estimation proposed in \cite{Tanaka1997a} is obtained as the lowest order of the Taylor series.
Using numerical simulations, we find that our estimation is more accurate for a much wider range of parameters compared to previously obtained estimations.
Therefore, the new estimation results to more accurate self-consistent equations for second-order oscillators.

Third, using the more accurate self-consistent equations, we have performed a detailed analysis of the properties of the steady state solutions, such as the existence of multiple branches, and their dependence on the initial state.
The critical transition point $K = K_c$ has also been calculated, coinciding with the stability analysis in \cite{Barre2016}, obtained for symmetric and unimodal distribution $g(\Omega)$  through an unstable manifold expansion. 

Finally, to better understand the dynamics and the  steady states, we have introduced new concepts such as the \emph{margin region}, Sec.~\ref{sec_margin_region}, and the \emph{scaled inertia} $\nu = s \mu$, Sec.~\ref{sec_scale_free_inertia}. 

The approach to self-consistent equations for second-order oscillators used in this paper provides a framework that can be easily generalized to explore properties of steady states for more general systems, for example, with non-constant inertias and damping coefficients or with phase shifts.
Moreover, combined with the development of generalized order parameters, as in \cite{Schroder2017universal}, our approach can also pave the way to the analysis of second-order oscillators in complex networks, such as power grids.
The analysis in this paper is from these points of view a basic building block in this research direction.

\section*{Acknowledgments}

We thank the Center for Information Technology of the University of Groningen for the use of the Peregrine HPC cluster for our numerical simulations.
We also thank the (anonymous) referees for their comments which helped to improve the presentation of this work.
J. Gao is supported by a China Scholarship Council (CSC) scholarship.

\bibliography{papersnew}

\end{document}